\documentclass[11pt]{article}
\usepackage{amsmath,amsfonts,amssymb}
\usepackage[all]{xy}
\usepackage{pstricks}
\numberwithin{equation}{section}

\newcommand{\nc}{\newcommand}
\nc\disp{\displaystyle}
\nc{\fh}{\hat{f}}
\nc{\muh}{\hat{\mu}}
\nc{\nuh}{\hat{\nu}}
\nc{\spos}[2]{\makebox(0,0)[#1]{$\sm{#2}$}}
\nc{\sm}[1]{{\scriptstyle #1}}
\nc{\bib}{\bibitem}
\nc{\al}{\alpha}
\nc{\g}{\gamma}
\nc{\G}{\Gamma}
\nc{\D}{\Delta}
\nc{\eps}{\epsilon}
\nc{\la}{\lambda}
\nc{\La}{\Lambda}
\nc{\var}{\varphi}
\nc{\pa}{\partial}
\nc{\nn}{\nonumber \\ }
\nc{\hf}{\frac{1}{2}}
\nc{\dz}{\frac{dz}{2\pi i}}
\nc{\bin}[2]{\left(\!\!\!\begin{array}{c} {#1}\\ {#2} \end{array}\!\!\!\right)}
\nc{\be}{\begin{equation}}
\nc{\ee}{\end{equation}}
\nc{\bea}{\begin{eqnarray}}
\nc{\eea}{\end{eqnarray}}
\nc{\bra}[1]{\langle {#1}|}
\nc{\ket}[1]{|{#1}\rangle}
\nc{\ketw}[1]{({#1})^{\phantom{a}}_{\Wc}}
\nc{\ketwast}[1]{({#1})^{\raisebox{0.25ex}{$\ast$}}_{\Wc}}
\nc{\chit}{\raisebox{0.25ex}{$\chi$}}
\nc{\chih}{\raisebox{0.25ex}{$\hat\chi$}}
\nc{\Db}{\mbox{\boldmath $D$}}
\nc{\Hb}{\mbox{\boldmath $H$}}
\nc{\calH}{{\cal H}}
\nc{\calR}{{\cal R}}
\nc{\calL}{{\cal L}}
\nc{\calV}{{\cal V}}
\nc{\Hc}{{\cal H}}
\nc{\Rc}{{\cal R}}
\nc{\Lc}{{\cal L}}
\nc{\Vc}{{\cal V}}
\nc{\Ib}{\mbox{\boldmath $I$}}
\nc{\qb}{\bar{q}}
\nc{\Ac}{\mathcal{A}}
\nc{\Bc}{\mathcal{B}}
\nc{\Cc}{\mathcal{C}}
\nc{\Dc}{\mathcal{D}}
\nc{\Ec}{\mathcal{E}}
\nc{\Fc}{\mathcal{F}}
\nc{\Gc}{\mathcal{G}}
\nc{\Ic}{\mathcal{I}}
\nc{\Jc}{\mathcal{J}}
\nc{\Kc}{\mathcal{K}}
\nc{\Mc}{\mathcal{M}}
\nc{\Nc}{\mathcal{N}}
\nc{\Oc}{\mathcal{O}}
\nc{\Pc}{\mathcal{P}}
\nc{\Qc}{\mathcal{Q}}
\nc{\Sc}{\mathcal{S}}
\nc{\Wc}{\mathcal{W}}
\nc{\Xc}{\mathcal{X}}
\nc{\Yc}{\mathcal{Y}}
\nc{\Zc}{\mathcal{Z}}
\nc{\fus}{\mbox{}\,\hat\otimes\,\mbox{}}
\nc{\Sh}{\hat{S}}
\def\vvdots{\mathinner{\mkern1mu\raise1pt\vbox{\kern7pt\hbox{.}}\mkern2mu
  \raise4pt\hbox{.}\mkern2mu\raise7pt\hbox{.}\mkern1mu}}
\nc{\gauss}[2]{\left[\!\!\begin{array}{c} {#1}\\ {#2} \end{array}\!\!\right]}
\nc{\sbin}[2]{\left\{\!\!\!\begin{array}{c} {#1}\\ {#2} 
\end{array}\!\!\!\right\}}
\nc{\sbinlr}[2]{\Big\langle\!\!\begin{array}{c} {#1}\\ {#2} 
\end{array}\!\!\Big\rangle}
\nc{\bino}[2]{\left(\!\!\begin{array}{c} {#1}\\ {#2} \end{array}\!\!\right)}
\definecolor{lightblue}{rgb}{.61,.61,1}
\definecolor{midblue}{rgb}{.7,.7,1}
\definecolor{lightlightblue}{rgb}{.85,.85,1}
\definecolor{lightestblue}{rgb}{.96,.96,1}
\definecolor{lightpurple}{rgb}{1,.65,1}
\nc{\ch}{{\rm ch}}
\nc{\R}{{\cal R}}
\nc{\dkk}{\delta_{j,\{k,k'\}}^{(2)}}
\nc{\drr}{\delta_{j,\{r,r'\}}^{(2)}}
\nc{\ddkk}{\delta_{j,\{k,k'\}}^{(4)}}
\nc{\dddkk}{\delta_{j,\{k,k'\}}^{(8)}}
\nc{\dnn}{\delta_{j,\{n,n'\}}^{(2)}}
\nc{\ddnn}{\delta_{j,\{n,n'\}}^{(4)}}
\nc{\dddnn}{\delta_{j,\{n,n'\}}^{(8)}}

\definecolor{pink}{rgb}{1,.65,.65}

\begin{document}

\topmargin -5mm
\oddsidemargin 5mm

\setcounter{page}{1}

\vspace{8mm}
\begin{center}
{\huge {\bf Fusion of irreducible modules in ${\cal WLM}(p,p')$}}

\vspace{8mm}
 {\LARGE J{\o}rgen Rasmussen}
\\[.3cm]
 {\em Department of Mathematics and Statistics, University of Melbourne}\\
 {\em Parkville, Victoria 3010, Australia}
\\[.4cm]
 {\tt j.rasmussen@ms.unimelb.edu.au}

\end{center}

\vspace{8mm}
\centerline{{\bf{Abstract}}}
\vskip.4cm
\noindent
Based on symmetry principles, we derive 
a fusion algebra generated from repeated fusions
of the irreducible modules appearing in the ${\cal W}$-extended logarithmic minimal model
${\cal WLM}(p,p')$. In addition to the irreducible 
modules themselves, closure of the commutative and associative fusion algebra requires 
the participation of a variety of reducible yet indecomposable modules.
We conjecture that this fusion algebra is the same as the one obtained
by application of the Nahm-Gaberdiel-Kausch algorithm and find that it reproduces
the known such results for ${\cal WLM}(1,p')$ and ${\cal WLM}(2,3)$.
For $p>1$, this fusion algebra does not contain a unit.
Requiring that the spectrum of modules is invariant under a natural notion of conjugation, 
however, introduces an additional
$(p-1)(p'-1)$ reducible yet indecomposable rank-1 modules, among which
the identity is found, still yielding a well-defined fusion algebra. In this greater fusion
algebra, the aforementioned symmetries are generated by fusions with the three irreducible
modules of conformal weights $\D_{kp-1,1}$, $k=1,2,3$.
We also identify polynomial fusion rings associated with our fusion algebras.
\renewcommand{\thefootnote}{\arabic{footnote}}
\setcounter{footnote}{0}

\section{Introduction}

We consider the infinite series of Yang-Baxter integrable logarithmic minimal models 
${\cal LM}(p,p')$~\cite{PRZ0607,RP0707} viewed in the ${\cal W}$-extended 
picture~\cite{PRR0803,RP0804,Ras0805,Ras0812} where they are denoted by
${\cal WLM}(p,p')$. The extension is believed to be
with respect to the $\Wc=\Wc_{p,p'}$ symmetry algebra of~\cite{FGST0606}, and we are
considering the models in their continuum scaling limits. 
An object of great interest is the fusion algebra, here denoted by $\mathrm{Irr}[{\cal WLM}(p,p')]$,
generated from repeated fusions of the $2pp'+\frac{1}{2}(p-1)(p'-1)$ 
irreducible modules in ${\cal WLM}(p,p')$. For $p>1$, we do not have boundary
conditions associated with all of these modules and are therefore incapable of
determining the complete set of fusion rules within our lattice approach based on
Cardy's picture~\cite{Car86,Car89}.
On the other hand, this fusion algebra is believed to be obtainable using the 
Nahm-Gaberdiel-Kausch algorithm~\cite{Nahm9402,GK9604}.
The application of this algorithm is
very tedious, however, even for $p=1$ and in the case ${\cal WLM}(2,3)$, 
so the goal here is to access the fusion rules by other means. 

First, we construct a fusion algebra whose spectrum of modules contains
all the irreducible modules in addition to 
$8pp'-6p-6p'+4$ reducible yet indecomposable modules of which 
$2(p-1)(p'-1)$ are of rank 1, $4pp'-2p-2p'$ are of rank 2, and $2(p-1)(p'-1)$ are of rank 3.
This fusion algebra is obtained from the fundamental fusion algebra 
$\mathrm{Fund}[{\cal WLM}(p,p')]$, defined in~\cite{Ras0812}, as the minimal extension
thereof which is invariant under a particular triplet of symmetries. 
Here, we say that a fusion algebra $\Ac$ with fusion multiplication $\otimes$
is invariant under $\Oc$, or simply $\Oc$-symmetric, if
\be 
 \Oc\big[\R\big]\otimes\R'=\R\otimes\Oc\big[\R'\big]=\Oc\big[\R\otimes\R'\big],
   \qquad\quad \forall\ \!\R,\R'\in\Jc
\label{ORR}
\ee
where $\Oc$ is a
map from and to the spectrum or set $\Jc$ of modules underlying the fusion algebra.
The spectrum of this extension of $\mathrm{Fund}[{\cal WLM}(p,p')]$ is also
invariant under a natural notion of conjugation. 
The extended fusion algebra itself is therefore denoted by 
$\mathrm{Conj}[{\cal WLM}(p,p')]$.
We prove that the three symmetries of $\mathrm{Conj}[{\cal WLM}(p,p')]$
are generated by fusion with the three irreducible
modules of conformal weights $\D_{kp-1,1}=\D_{1,kp'-1}$, $k=1,2,3$, 
where $\D_{\rho,\sigma}$ is given by the usual Kac formula.
For critical percolation in the ${\cal W}$-extended picture ${\cal WLM}(2,3)$,
these conformal weights are $\D_{1,1}=0$, $\D_{3,1}=2$ and $\D_{5,1}=7$.

As a subalgebra of $\mathrm{Conj}[{\cal WLM}(p,p')]$, we identify the algebra
generated by repeated fusions of the irreducible modules.
We conjecture that this fusion algebra is indeed $\mathrm{Irr}[{\cal WLM}(p,p')]$
and note that it is obtained from $\mathrm{Conj}[{\cal WLM}(p,p')]$
by omitting $(p-1)(p'-1)$ of the reducible yet indecomposable rank-1 modules.
That our proposal for ${\cal WLM}(1,p')$ yields the known 
results~\cite{GK9606,GR0707,PRR0803} is ensured by construction,
while we have verified that it also reproduces the very recent results for
${\cal WLM}(2,3)$~\cite{GRW0905}.

For $p>1$, the fusion algebra $\mathrm{Irr}[{\cal WLM}(p,p')]$ does {\em not} have a unit
nor is its spectrum invariant under conjugation.
The minimal extension, whose spectrum {\em is} conjugation invariant, is
the fusion algebra $\mathrm{Conj}[{\cal WLM}(p,p')]$,
and this algebra {\em does} contain an identity. In this setting, conjugation invariance
of the spectrum thus implies the existence of an identity.
The paper~\cite{GRW0905} on ${\cal WLM}(2,3)$ is actually focussed on such a 
conjugation-invariant spectrum, and we
have verified that $\mathrm{Conj}[{\cal WLM}(2,3)]$ indeed corresponds to their results.

We also identify a polynomial fusion ring isomorphic to 
$\mathrm{Conj}[{\cal WLM}(p,p')]$. 
For $p=1$, where $\mathrm{Conj}[{\cal WLM}(1,p')]=\mathrm{Fund}[{\cal WLM}(1,p')]$, 
this was already done in~\cite{Ras0812} and involved a quotient polynomial ring
with two generators. For $p>1$, on the other hand, we find that
the sought-after quotient polynomial ring has {\em five} generators corresponding
to the two fundamental representations of 
$\mathrm{Fund}[{\cal WLM}(p,p')]$ and the three symmetry-generating irreducible
modules $\Wc(\D_{kp-1,1})$, $k=1,2,3$.
The fusion algebra $\mathrm{Irr}[{\cal WLM}(p,p')]$ is isomorphic to a subring
thereof.

Many subalgebras and quotients can be identified of the various fusion algebras discussed here. 
In~\cite{Ras0908}, a general framework is outlined within which it makes sense to discuss 
rings of equivalence classes of fusion-algebra generators. Grothendieck-like rings, as the one 
generated by the $2pp'$ generators $\Kc^{\pm}_{r,s}$ in~\cite{FGST0606}, arise as 
particularly interesting cases obtained by elevating character identities to 
equivalence relations between the corresponding fusion generators.

\subsection*{Notation and terminology}
\vskip.1cm 
Unless otherwise specified, we let
\be
 \kappa,\kappa'\in\mathbb{Z}_{1,2},\quad r\in\mathbb{Z}_{1,p},\quad
  s\in\mathbb{Z}_{1,p'},\quad a,a'\in\mathbb{Z}_{1,p-1},\quad b,b'\in\mathbb{Z}_{1,p'-1},
  \quad\al\in\mathbb{Z}_{0,p-1},\quad\beta\in\mathbb{Z}_{0,p'-1}
\label{labels}
\ee
where
\be
 \mathbb{Z}_{n,m}=\mathbb{Z}\cap[n,m],\qquad\quad n,m\in\mathbb{Z}
\ee 
denotes the set of integers from $n$ to $m$, both included. 
By an expression like $\kappa\cdot\kappa'$, we mean 
$1\cdot1=2\cdot2=1$ or $1\cdot2=2\cdot1=2$. As a simplified notation for 
a set of elements with labels of the form (\ref{labels}), we write
\be
 \big\{f_{\kappa,a},g_{\alpha}^{0,s}\big\}
  =\big\{f_{\kappa,a};\ \kappa\in\mathbb{Z}_{1,2},a\in\mathbb{Z}_{1,p-1}\big\}
  \cup\big\{g_{\alpha}^{0,s};\ \alpha\in\mathbb{Z}_{0,p-1},s\in\mathbb{Z}_{1,p'}\big\}
\ee
for example.
The two terms {\em representation} and {\em module} are often used interchangeably when
the discussion is on modules. Here, we use the term module.

\subsection*{Sets of indecomposable modules and their intersection diagram}

To assist the reader, the various sets of indecomposable modules are summarized here.
Their intersection diagram appears in (\ref{int}) below. As convenient abbreviations, in (\ref{int}),
we let $\ketw{\R_2}$ and $\ketw{\R_3}$ denote the sets of indecomposable 
modules of rank 2 and 3, respectively, while $\Wc(\D_{p-a,b})$ represents the set 
$\{\Wc(\D_{p-a,b});\ a\in\mathbb{Z}_{1,p-1},\ b\in\mathbb{Z}_{1,p'-1}\}$, and so on.
If $\Jc$ is a subset of $\Jc'$, the fusion algebra $\langle\Jc\rangle$ generated from 
$\Jc$ is a subalgebra of the fusion algebra $\langle\Jc'\rangle$ generated from $\Jc'$.
It is noted that the fusions may generate indecomposable modules not listed explicitly
in the two sets.

Now, the set of indecomposable modules associated with boundary conditions is denoted by 
$\ketw{\Jc_{p,p'}^{\mathrm{out}}}$ and is represented in (\ref{int}) by the interior of the hexagon.
The set of indecomposable modules appearing in the fundamental fusion algebra
is denoted by $\ketw{\Jc_{p,p'}^{\mathrm{fund}}}$ 
and is represented by the interior of the upward-pointing triangle.
The set of irreducible modules is denoted by $\Jc_{p,p'}^{\mathrm{irr}}$ and is represented 
by the interior of the soft-cornered square.
The set of indecomposable modules generated by repeated fusions of the irreducible modules
is denoted by $\ketw{\Jc_{p,p'}^{\mathrm{irr}}}$ and is represented by the interior of the 
downward-pointing triangle.
The total set of indecomposable modules considered here is denoted by 
$\ketw{\Jc_{p,p'}^{\mathrm{conj}}}$.
The cardinalities of these sets are
\bea
 &\big|\ketw{\Jc^{\mathrm{out}}_{p,p'}}\big|\;=\;6pp'-2p-2p',\qquad
    \big|\ketw{\Jc^{\mathrm{fund}}_{p,p'}}\big|\;=\;7pp'-3p-3p'+1\nn
 & \big|J^{\mathrm{irr}}_{p,p'}\big|\;=\;2pp'+\frac{1}{2}(p-1)(p'-1),\qquad
    \big|\ketw{\Jc^{\mathrm{irr}}_{p,p'}}\big|\;=\;4p+4p'-6+\frac{19}{2}(p-1)(p'-1)\nn
 &\big|\ketw{\Jc^{\mathrm{conj}}_{p,p'}}\big|
    \;=\;4p+4p'-6+\frac{21}{2}(p-1)(p'-1)
\eea
\be
 \mbox{
 \begin{picture}(100,320)(0,-150)
    \unitlength=1.5cm
  \thinlines
\put(-2,0){\line(1,0){6}}
\put(-2,0){\line(3,4){3}}
\put(4,0){\line(-3,4){3}}
\put(-2.5,2.5){\line(1,0){7}}
\put(-2.5,2.5){\line(3,-5){3.5}}
\put(4.5,2.5){\line(-3,-5){3.5}}
\put(0.2,0.1){\line(1,0){1.6}}
\put(0.2,0.1){\line(-1,1){1.15}}
\put(0.2,2.4){\line(-1,-1){1.15}}
\put(0.2,2.4){\line(1,0){1.6}}
\put(1.8,0.1){\line(1,1){1.15}}
\put(1.8,2.4){\line(1,-1){1.15}}
\put(1,0){\oval(3,3)}
\put(0.65,3){$\ketw{a,b}$}
\put(0.15,1.9){$\ketw{\R_2}$}
\put(1.15,1.9){$\ketw{\R_3}$}
\put(-0.25,1){$\Wc(\D_{\kappa p,b})$}
\put(1.15,1){$\Wc(\D_{a,\kappa p'})$}
\put(0.45,0.3){$\Wc(\D_{\kappa p,p'})$}
\put(-0.35,-0.5){$\Wc(\D_{p-a,b})$}
\put(1.05,-0.5){$\Wc(\D_{2p-a,b})$}
\put(0.35,-1.2){$\Wc(\D_{3p-a,b})$}
\put(0.65,-2.2){$\ketwast{a,b}$}
\put(4.05,-0.05){$\ketw{\Jc_{p,p'}^{\mathrm{fund}}}$}
\put(4.55,2.4){$\ketw{\Jc_{p,p'}^{\mathrm{irr}}}$}
\put(3,1.2){$\leftarrow----$}
\put(4.3,1.2){$\ketw{\Jc_{p,p'}^{\mathrm{out}}}$}
\put(2.45,-1.45){$\Jc_{p,p'}^{\mathrm{irr}}$}
\put(2.8,3.2){$\ketw{\Jc_{p,p'}^{\mathrm{conj}}}$}
 \end{picture}
}
\label{int}
\ee

\section{Fundamental fusion algebra of ${\cal WLM}(p,p')$}

A logarithmic minimal model ${\cal LM}(p,p')$ is defined~\cite{PRZ0607,RP0707} 
for every coprime pair of positive integers $p<p'$. The model has central charge
\be
 c\ =\  1-6\frac{(p'-p)^2}{pp'}
\label{c}
\ee
and conformal weights
\be
 \D_{\rho,\sigma}\ =\ \frac{(\rho p'-\sigma p)^{2}-(p'-p)^2}{4pp'},\hspace{1.2cm} 
  \rho,\sigma\in\mathbb{N}
\label{D}
\ee
Its ${\cal W}$-extension ${\cal WLM}(p,p')$ is discussed 
in~\cite{PRR0803,RP0804,Ras0805,Ras0812} and briefly reviewed in the following.

\subsection{Modules associated with boundary conditions}

The indecomposable modules in ${\cal WLM}(p,p')$ 
which can be associated with Yang-Baxter integrable boundary conditions
on the strip lattice and ${\cal W}$-invariant
boundary conditions in the continuum scaling limit,
were identified in~\cite{RP0804,Ras0805} by extending constructions in~\cite{PRR0803}
pertaining to the case $p=1$. The set of these modules is given by
\be
 \ketw{\Jc^{\mathrm{out}}_{p,p'}}=\big\{
  \Wc(\D_{\kappa p,b}),\Wc(\D_{a,\kappa p'}),\Wc(\D_{\kappa p,p'}),
   \ketw{\R_{\kappa p,s}^{a,0}},\ketw{\R_{r,\kappa p'}^{0,b}},\ketw{\R_{\kappa p,p'}^{a,b}}\big\}
\label{JWout}
\ee
and is of cardinality
\be
 \big|\ketw{\Jc^{\mathrm{out}}_{p,p'}}\big|\ =\ 6pp'-2p-2p'
\ee
Here, we have adopted the notation of~\cite{GRW0905} denoting a
${\cal W}$-irreducible module of conformal weight $\D$ by $\Wc(\D)$.
Thus, there are $2p+2p'-2$ irreducible (hence indecomposable rank-1) modules
\be
 \big\{\Wc(\D_{\kappa p,s}),\Wc(\D_{r,\kappa p'})\big\}
\label{r1}
\ee
where the two modules $\Wc(\D_{\kappa p,p})=\Wc(\D_{p,\kappa p'})$ are listed twice,
in addition to $4pp'-2p-2p'$ indecomposable rank-2 modules
\be
 \big\{\ketw{\R_{\kappa p,s}^{a,0}}, \ketw{\R_{r,\kappa p'}^{0,b}}\big\}
\label{r2}
\ee
and $2(p-1)(p'-1)$ indecomposable rank-3 modules
\be
 \big\{\ketw{\R_{\kappa p,\kappa' p'}^{a,b}}\big\}
 \hspace{1.2cm}\mathrm{subject\ to}\ \ \ 
  \ketw{\R_{p,2p'}^{a,b}}\equiv\ketw{\R_{2p,p'}^{a,b}}\quad
   \mathrm{and}\quad \ketw{\R_{2p,2p'}^{a,b}}\equiv\ketw{\R_{p,p'}^{a,b}}
\label{r3}
\ee
The fusion algebra of these modules~\cite{Ras0805,Ras0812}
\be
 \mathrm{Out}[{\cal WLM}(p,p')]=
  \big\langle \ketw{\Jc^{\mathrm{out}}_{p,p'}}\big\rangle
\label{WfusOut}
\ee
is given explicitly in Appendix~\ref{AppFus} as (\ref{fus11}) through (\ref{fus33})
and is both associative and commutative. There is no unit or identity for $p>1$, while, for
$p=1$, the irreducible module $\Wc(\D_{1,1})$ is the identity.

In~\cite{Ras0805}, it was conjectured that every indecomposable rank-2 module
has an embedding pattern of one of the types 
\be
 \mbox{
 \begin{picture}(100,120)(0,0)
    \unitlength=1cm
  \thinlines
\put(-2.3,2){$\mathcal{E}(\D_h,\D_v):$}
\put(1.9,3.6){$\Wc(\D_v)$}
\put(-0.1,2){$\Wc(\D_h)$}
\put(3.9,2){$\Wc(\D_h)$}
\put(1.9,0.4){$\Wc(\D_v)$}
\put(3.5,2.1){\vector(-1,0){2.1}}
\put(1.9,3.4){\vector(-4,-3){1.2}}
\put(4.1,1.7){\vector(-4,-3){1.2}}
\put(1.9,0.8){\vector(-4,3){1.2}}
\put(4.1,2.5){\vector(-4,3){1.2}}
 \end{picture}
}
\hspace{5.3cm}
 \mbox{
 \begin{picture}(100,120)(-10,0)
    \unitlength=1cm
  \thinlines
\put(-3,2){$\mathcal{E}(\D_h,\D_v;\D_c):$}
\put(1.9,3.6){$\Wc(\D_v)$}
\put(-0.1,2){$\Wc(\D_h)$}
\put(3.9,2){$\Wc(\D_h)$}
\put(1.9,0.4){$\Wc(\D_v)$}
\put(3.5,2.1){\vector(-1,0){2.1}}
\put(1.9,3.4){\vector(-4,-3){1.2}}
\put(4.1,1.7){\vector(-4,-3){1.2}}
\put(1.9,0.8){\vector(-4,3){1.2}}
\put(4.1,2.5){\vector(-4,3){1.2}}
\put(1.9,1.35){$\Wc(\D_c)$}
\put(3.7,1.9){\vector(-2,-1){0.6}}
\put(1.8,1.6){\vector(-2,1){0.5}}
 \end{picture}
}
\label{E}
\ee 
where the horizontal arrows indicate the non-diagonal action of the Virasoro mode $L_0$.
Specifically, the indecomposable rank-2 modules 
(\ref{r2}) are believed to enjoy the embedding patterns 
\bea
  &&\ketw{\R_{p,b}^{a,0}}\sim\Ec(\D_{p+a,b},\D_{3p-a,b};\D_{p-a,b}),\qquad\quad\ \
    \ketw{\R_{a,p'}^{0,b}}\sim\Ec(\D_{a,p'+b},\D_{a,3p'-b};\D_{a,p'-b})\nn
  &&\ketw{\R_{p,p'}^{a,0}}\sim\Ec(\D_{p+a,p'},\D_{3p-a,p'}),\qquad\qquad\qquad
    \ketw{\R_{p,p'}^{0,b}}\sim\Ec(\D_{p,p'+b},\D_{p,3p'-b})\nn
  &&\ketw{\R_{2p,s}^{a,0}}\sim\Ec(\D_{2p+a,s},\D_{2p-a,s}),\qquad\qquad\qquad
    \ketw{\R_{r,2p'}^{0,b}}\sim\Ec(\D_{r,2p'+b},\D_{r,2p'-b})
\label{Rr2E}
\eea
In~\cite{Ras0805}, it was also conjectured that the
indecomposable rank-3 modules (\ref{r3}) have embedding structures
described by the patterns in (\ref{E}), namely
\be
 \ketw{\R_{\kappa p,p'}^{a,b}}
  \sim\Ec\Big(\ketw{\R_{\kappa p,p'-b}^{a,0}},\ketw{\R_{(3-\kappa)p,b}^{a,0}}\Big)
  \sim\Ec\Big(\ketw{\R_{p-a,\kappa p'}^{0,b}},\ketw{\R_{a,(3-\kappa)p'}^{0,b}}\Big)
\label{Rr3E}
\ee
where the irreducible modules $\Wc(\D_h)$ and $\Wc(\D_v)$
have been replaced by indecomposable rank-2 modules.

\subsection{Supplementary modules}

In~\cite{Ras0812}, based on algebraic arguments,
we suggested to supplement the set of indecomposable
modules (\ref{JWout}) by the reducible yet indecomposable rank-1 modules 
\be
 \big\{\ketw{a,b}\big\}
\label{ab}
\ee
The cardinality of the disjoint union
\be
 \ketw{\Jc^{\mathrm{fund}}_{p,p'}}=\big\{\ketw{a,b}\big\}\cup\ketw{\Jc^{\mathrm{out}}_{p,p'}}
\label{JWfund}
\ee
is therefore given by
\be
 \big|\ketw{\Jc^{\mathrm{fund}}_{p,p'}}\big|=
  (p-1)(p'-1)+|\ketw{\Jc^{\mathrm{out}}_{p,p'}}|=7pp'-3p-3p'+1
\label{CardJFund}
\ee
We also argued that the embedding pattern of $\ketw{a,b}$ is of the form
\be
 \mbox{
 \begin{picture}(100,60)(-10,0)
    \unitlength=0.8cm
  \thinlines
\put(-0.7,1.9){$\Wc(\D_{2p-a,b})$}
\put(2.7,0){$\Wc(\D_{a,b})$}
\put(2.6,0.6){\vector(-4,3){1.2}}
 \end{picture}
}
\label{abemb}
\ee
implying the short exact sequence
\be
 0\to\Wc(\D_{2p-a,b})\to\ketw{a,b}\to\Wc(\D_{a,b})\to0
\ee

The algebraic extension (\ref{JWfund}) of the set of indecomposable modules (\ref{JWout}) was 
shown in~\cite{Ras0812} to yield a well-defined fusion algebra called the
fundamental fusion algebra and denoted by
\be
 \mathrm{Fund}[{\cal WLM}(p,p')]=
  \big\langle \ketw{\Jc^{\mathrm{fund}}_{p,p'}}\big\rangle
\label{WfusFund}
\ee
The underlying fusion rules are all listed in Appendix~\ref{AppFus}.
The algebra is generated from repeated fusions of the two `fundamental representations' 
$\ketw{2,1}$ and $\ketw{1,2}$ (strictly speaking, in addition to the identity $\ketw{1,1}$)
\be
 \mathrm{Fund}[{\cal WLM}(p,p')]=\big\langle \ketw{1,1}, \ketw{2,1},\ketw{1,2}\big\rangle
\label{Fund2112}
\ee
{}From~\cite{Ras0812}, based on an explicit inspection of the fusion rules,
we know that $\mathrm{Out}[{\cal WLM}(p,p')]$ is an ideal of 
$\mathrm{Fund}[{\cal WLM}(p,p')]$. Since the set (\ref{ab}) is empty for $p=1$, we note that 
\be
 \mathrm{Fund}[{\cal WLM}(1,p')]=\mathrm{Out}[{\cal WLM}(1,p')]
\ee

\section{Fusion of irreducible modules in ${\cal WLM}(p,p')$}

\subsection{Modules}

It is the same set of irreducible modules which appears
as subfactors of the indecomposable modules in (\ref{JWout}) as in (\ref{JWfund}).
This set is given by
\be
 J^{\mathrm{irr}}_{p,p'}=\big\{\Wc(\D_{\rho,\sigma});\ \rho p'\geq \sigma p,
   \rho\in\mathbb{Z}_{1,3p-1},\sigma\in\mathbb{Z}_{1,p'}\big\}
\label{irr}
\ee
and we recall the simple identities
\be
 \D_{a,b}=\D_{a+kp,b+kp'},\qquad \D_{a,kp'-b}=\D_{kp-a,b},\qquad\quad k\in\mathbb{Z}
\ee
allowing a great deal of freedom in the labeling of the conformal weights.
As a matter of convention, we have chosen the labeling indicated in (\ref{irr}).
This set of irreducible modules also appears in~\cite{FGST0606} and has cardinality
\be
 \big|J^{\mathrm{irr}}_{p,p'}\big|=2pp'+\frac{1}{2}(p-1)(p'-1)
\ee
For $p>1$, the set (\ref{irr}) is larger than the set of
irreducible modules (\ref{r1}) appearing as generators in the fundamental fusion algebra.
It is thus natural to try to understand the fusion algebra
resulting from repeated fusions of the irreducible modules (\ref{irr}) and
to determine the set of modules required to ensure closure
of this fusion algebra.
Even though the set (\ref{r1}) is a subset of (\ref{irr}),
there is, a priori, no need for the fundamental fusion algebra to be
a subalgebra of this fusion algebra since the former is generated
by the two fundamental representations which {\em may not}, after all, arise from
repeated fusions of the irreducible modules in (\ref{irr}).  
Indeed, the fusion algebra $\mathrm{Irr}[{\cal WLM}(p,p')]$, to be discussed below, does {\em not}
contain the fundamental fusion algebra as a subalgebra, while
the fusion algebra $\mathrm{Conj}[{\cal WLM}(p,p')]$, also to be discussed below, {\em does}.

A complicating factor for $p>1$ is that we do not have boundary
conditions associated with all of the irreducible modules in (\ref{irr})
(only with the ones appearing in (\ref{r1})) and are therefore incapable of
determining the complete set of fusion rules within our lattice approach.
On the other hand, the sought-after fusion algebra is believed to be obtainable using the 
Nahm-Gaberdiel-Kausch algorithm~\cite{Nahm9402,GK9604}.
The application of this algorithm is
very tedious, however, even for $p=1$ and in the case ${\cal WLM}(2,3)$, 
so an alternative approach to the fusion rules is certainly welcome. 
Our proposal below is to use symmetry principles, and we have verified, as we will discuss, 
that our conjectured fusion algebras indeed 
reproduce the known results obtained using the algorithm.

In preparation for the discussion of fusion rules, we introduce the $(p-1)(p'-1)$
reducible yet indecomposable rank-1 modules
\be
 \big\{\ketwast{a,b}\big\}
\label{abast}
\ee
whose embedding patterns 
\be
 \mbox{
 \begin{picture}(100,60)(-10,0)
    \unitlength=0.8cm
  \thinlines
\put(-0.7,1.9){$\Wc(\D_{2p-a,b})$}
\put(2.7,0){$\Wc(\D_{a,b})$}
\put(1.5,1.5){\vector(4,-3){1.2}}
 \end{picture}
}
\label{abastemb}
\ee
imply the short exact sequences
\be
 0\to\Wc(\D_{a,b})\to\ketwast{a,b}\to\Wc(\D_{2p-a,b})\to0
\ee
It follows immediately that 
\be
 \chit[\ketwast{a,b}](q)=\chit[\ketw{a,b}](q)
\ee
where the characters $\chit[\ketw{a,b}](q)$ are discussed in~\cite{Ras0812}
alongside the characters of all the other modules appearing in the fusion algebra.

It may seem surprising that we are introducing the contragredient modules $\ketwast{a,b}$.
To motivate their appearance, we briefly consider fusion of the underlying Virasoro 
modules in ${\cal LM}(p,p')$. Details thereof may be found in~\cite{EF0604},
in particular in the case ${\cal LM}(2,3)$, and are obtained using the Nahm-Gaberdiel-Kausch
algorithm. 
Let us denote by $\Vc(\D)$ the irreducible Virasoro module of conformal weight $\D$.
A careful re-examination of the fusion $\Vc(2)\otimes\Vc(2)$ in ${\cal LM}(2,3)$
reveals that a natural but incorrect identification was made in~\cite{EF0604}. 
This was also observed in~\cite{GRW0905}. The correct fusion rule reads
\be
  \Vc(2)\otimes\Vc(2)=(1,1)^{\raisebox{0.25ex}{$\ast$}}
\label{22}
\ee
where $(1,1)^{\raisebox{0.25ex}{$\ast$}}$ is the indecomposable module
contragredient to the indecomposable identity module $(1,1)$
\be
 0\to\Vc(2)\to(1,1)\to\Vc(0)\to0,\qquad\quad
 0\to\Vc(0)\to(1,1)^{\raisebox{0.25ex}{$\ast$}}\to\Vc(2)\to0
\ee
Continuing this analysis also sees the introduction of the two indecomposable Virasoro
modules $(1,2)$ and $(1,2)^{\raisebox{0.25ex}{$\ast$}}$ corresponding to the 
short exact sequences
\be
 0\to\Vc(1)\to(1,2)\to\Vc(0)\to0,\qquad\quad
 0\to\Vc(0)\to(1,2)^{\raisebox{0.25ex}{$\ast$}}\to\Vc(1)\to0
\ee
In ${\cal LM}(p,p')$, this generalizes to pairs $(a,b)$ and $(a,b)^{\raisebox{0.25ex}{$\ast$}}$
whose ${\cal W}$-extended counterparts in ${\cal WLM}(p,p')$
we have denoted by $\ketw{a,b}$ and $\ketwast{a,b}$, respectively.

Returning to the preparations, we also introduce the sets
\be
 \ketw{\Jc^{\mathrm{conj}}_{p,p'}}=\ketw{\Jc^{\mathrm{irr}}_{p,p'}}\cup\big\{\ketw{a,b}\big\},
     \qquad
  \ketw{\Jc^{\mathrm{irr}}_{p,p'}}=J^{\mathrm{irr}}_{p,p'}\cup\big\{\ketwast{a,b},
    \ketw{\R_{\kappa p,s}^{a,0}}, \ketw{\R_{r,\kappa p'}^{0,b}},\ketw{\R_{\kappa p,p'}^{a,b}}\big\}
\ee
as disjoint unions. Their cardinalities are thus
\be
 \big|\ketw{\Jc^{\mathrm{conj}}_{p,p'}}\big|
   =\big|\ketw{\Jc^{\mathrm{irr}}_{p,p'}}\big|+(p-1)(p'-1),\qquad
 \big|\ketw{\Jc^{\mathrm{irr}}_{p,p'}}\big|=4p+4p'-6+\frac{19}{2}(p-1)(p'-1)
\label{JWirr}
\ee
and will appear as the dimensions of two of the fusion algebras to be discussed.
The notation $\ketw{\Jc^{\mathrm{conj}}_{p,p'}}$ 
and $\ketw{\Jc^{\mathrm{irr}}_{p,p'}}$ will become clear in the following.

\subsection{Spectrum maps}

To facilitate the description of the fusion algebra generated from repeated fusions
of the irreducible modules (\ref{irr}), we now introduce some maps from
$\ketw{\Jc^{\mathrm{conj}}_{p,p'}}$ to itself.
We first extend the use of $\ast$ in (\ref{abast}) to an involution, here denoted by $\Cc$ and
referred to as {\em conjugation}, on the entire set of 
modules $\ketw{\Jc^{\mathrm{conj}}_{p,p'}}$ by
\be
 \Cc\big[\ketw{a,b}\big]=\ketwast{a,b},\qquad \Cc\big[\ketwast{a,b}\big]=\ketw{a,b},\qquad
  \Cc\big[\R\big]=\R,\qquad 
   \R\in\ketw{\Jc^{\mathrm{out}}_{p,p'}}\cup J^{\mathrm{irr}}_{p,p'}
\label{ast}
\ee
Since the embedding patterns (\ref{Rr2E}) and (\ref{Rr3E})
are invariant under reversal of the arrows, we see that the conjugation
$\Cc$, as an operation on the
embedding patterns, simply reverses the arrows. We note that 
this is trivially true when applied also to the irreducible modules.

We also introduce the map $\Kc$ which, on $\ketw{\Jc^{\mathrm{out}}_{p,p'}}$,
acts by $\kappa\leftrightarrow2\cdot\kappa=3-\kappa$ on the labeling of the modules
as given in (\ref{JWout}), vanishes on $\{\Wc(\D_{a,b})\}$, while its action on
$\{\ketw{a,b},\ketwast{a,b},\Wc(\D_{\kappa p+a,b})\}$
is described by the diagram
\be
 \mbox{
 \begin{picture}(250,85)(50,0)
    \unitlength=0.9cm
  \thinlines
\put(2.5,0){$\ketw{a,b}$}
\put(5.9,0){$\Wc(\D_{3p-a,b})$}
\put(5.9,2.9){$\Wc(\D_{2p-a,b})$}
\put(10.3,0){$\ketwast{a,b}$}
\put(4,0.1){\vector(1,0){1.65}}
\put(9.5,0.1){\vector(-1,0){1.15}}
\put(9.5,0.1){\vector(1,0){.5}}
\put(7,2.5){\vector(0,-1){1.9}}
\put(4.65,.3){$\Kc$}
\put(9.05,.3){$\Kc$}
\put(7.15,1.4){$\Kc$}
 \end{picture}
}
\label{symmKc}
\ee
We thus have
\bea
 &&\Kc\big[\ketw{\R_{\kappa p,s}^{a,0}}\big]=\ketw{\R_{(2\cdot\kappa)p,s}^{a,0}},\qquad
  \Kc\big[\ketw{\R_{r,\kappa p'}^{0,b}}\big]=\ketw{\R_{r,(2\cdot\kappa)p'}^{0,b}},\qquad
  \Kc\big[\ketw{\R_{\kappa p,p'}^{a,b}}\big]=\ketw{\R_{(2\cdot\kappa)p,p'}^{a,b}}\nn
 &&\Kc\big[\Wc(\D_{\kappa p,s})\big]=\Wc(\D_{(2\cdot\kappa)p,s}),\qquad
  \Kc\big[\Wc(\D_{\kappa p+a,p'})\big]=\Wc(\D_{(2\cdot\kappa)p+a,p'}),\qquad
  \Kc\big[\Wc(\D_{a,b})\big]=0  \nn
 &&\Kc\big[\ketw{a,b}\big]=\Kc\big[\ketwast{a,b}\big]
     =\Kc\big[\Wc(\D_{2p-a,b})\big]=\Wc(\D_{3p-a,b}),\qquad
  \Kc\big[\Wc(\D_{3p-a,b})\big]=\ketwast{a,b}
\label{Kc}
\eea
{\bf Lemma 1}\ \ \ The fusion algebra $\big\langle\ketw{\Jc^{\mathrm{out}}_{p,p'}}\big\rangle$
is $\Kc$-symmetric (in the sense of (\ref{ORR}) with $\Oc=\Kc$ restricted to
$\Jc=\ketw{\Jc^{\mathrm{out}}_{p,p'}}$).
\\[.2cm]
{\bf Lemma 2}\ \ \ In the fusion algebra $\big\langle\ketw{\Jc^{\mathrm{fund}}_{p,p'}}\big\rangle$,
we have
\be 
 \ketw{a,b}\otimes\Kc\big[\Qc\big]=\Kc\big[\ketw{a,b}\otimes\Qc\big],\qquad\quad
  \Qc\in\ketw{\Jc^{\mathrm{out}}_{p,p'}}
\label{abKQ}
\ee
Lemma 1 follows by direct inspection of the fusion rules (\ref{fus11}) though (\ref{fus33}),
while Lemma 2 follows by direct inspection of the fusion rules (\ref{sup1}) through (\ref{sup3}).

The map $\Lc$ is defined by
\bea
 &&\Lc\big[\ketw{a,b}\big]=\Wc(\D_{2p-a,b}),\qquad
   \Lc\big[\Wc(\D_{2p-a,b})\big]=\ketwast{a,b},\qquad
   \Lc\big[\Wc(\D_{a,b})\big]=0\nn
 &&\Lc\big[\R\big]=\R,\qquad \quad
   \R\in\ketw{\Jc^{\mathrm{irr}}_{p,p'}}\setminus\big\{\Wc(\D_{\kappa p-a,b})\big\}
\label{Lc}
\eea
while the map $\Mc$ is defined by
\bea
 &&\Mc\big[\ketw{a,b}\big]=\Wc(\D_{p-a,b}),\qquad
   \Mc\big[\Wc(\D_{a,b})\big]=\Wc(\D_{p-a,b})\nn
 &&\Mc\big[\R\big]=0,\qquad\quad
    \R\in\ketw{\Jc^{\mathrm{irr}}_{p,p'}}\setminus\big\{\Wc(\D_{a,b})\big\}
\label{Mc}
\eea
Since $\Lc$ and $\Mc$
both act trivially on $\ketw{\Jc^{\mathrm{out}}_{p,p'}}$, Lemma 1 and 2 obviously
apply also when replacing $\Kc$ by either $\Lc$ or $\Mc$ (recalling that 
$\ketw{\Jc^{\mathrm{out}}_{p,p'}}$ generates an ideal of
$\big\langle\ketw{\Jc^{\mathrm{fund}}_{p,p'}}\big\rangle$).
We note that the introduction of $\Lc$ and $\Mc$ is meaningless for $p=1$. 

Under composition, the maps $\Kc$, $\Lc$ and $\Mc$ generate a five-dimensional
commutative algebra whose composition rules in the basis
\be
 \Bc=\big\{\Kc,\Lc,\Lc^2,\Mc,\Mc^2\big\},\qquad\quad \Lc^2=\Lc\circ\Lc,\quad \Mc^2=\Mc\circ\Mc
\label{five}
\ee
are summarized in Figure~\ref{Cayley}.   
\begin{figure}
$$
\renewcommand{\arraystretch}{1.5}
\begin{array}{c||ccccc}
\circ&\Kc&\Lc&\Lc^2&\Mc&\Mc^2
\\[4pt]
\hline \hline
\rule{0pt}{14pt}
 \Kc
   &\Lc^2&\Kc&\Kc&0&0
\\[4pt]
 \Lc
   &\Kc&\Lc^2&\Lc^2&0&0
\\[4pt]
 \Lc^2
   &\Kc&\Lc^2&\Lc^2&0&0
\\[4pt]
 \Mc
   &0&0&0&\Mc^2&\Mc
\\[4pt]
 \Mc^2
   &0&0&0&\Mc&\Mc^2
\end{array}
$$
\caption{Cayley table of the composition rules for $\Kc,\Lc,\Lc^2,\Mc,\Mc^2\in\Bc$ for $p>1$.}
\label{Cayley}
\end{figure}
This algebra has no unit but can, of course, be extended straightforwardly by inclusion
of the identity map $\Ic$ on $\ketw{\Jc^{\mathrm{conj}}_{p,p'}}$.
We note that $\{\Mc,\Mc^2\}$, for example, generates an ideal. 
Another interesting observation is that $\Lc$ {\em itself} does not appear
as the result of composing any of the maps $\Kc,\Lc^2=\Kc\circ\Kc,\Mc,\Mc^2$.
We will return to this point in Section~\ref{SecProp}.

In partial summary, and with $\Qc_{\kappa}$ denoting a general element of 
$\ketw{\Jc^{\mathrm{out}}_{p,p'}}$, the diagrams
\be
 \mbox{
 \begin{picture}(100,120)(90,-30)
    \unitlength=0.9cm
  \thinlines
\put(-1.7,0){$\Wc(\D_{p-a,b})$}
\put(2.5,0){$\ketw{a,b}$}
\put(5.9,0){$\Wc(\D_{3p-a,b})$}
\put(5.9,2.9){$\Wc(\D_{2p-a,b})$}
\put(10.3,0){$\ketwast{a,b}$}
\put(2.2,0.1){\vector(-1,0){1.65}}
\put(4,0.1){\vector(1,0){1.65}}
\put(9.5,0.1){\vector(-1,0){1.15}}
\put(9.5,0.1){\vector(1,0){.5}}
\put(7,2.5){\vector(0,-1){1.9}}
\put(3.9,.55){\vector(1,1){1.9}}
\put(8.2,2.5){\vector(1,-1){1.9}}
\put(7,-0.7){\line(1,0){2.8}}
\put(9.8,-0.7){\vector(3,2){0.5}}
\put(7,-0.7){\line(-1,0){2.8}}
\put(4.2,-0.7){\vector(-3,2){0.5}}
\put(1.2,.3){$\Mc$}
\put(4.65,.3){$\Kc$}
\put(9.05,.3){$\Kc$}
\put(7.15,1.4){$\Kc$}
\put(4.5,1.7){$\Lc$}
\put(9.3,1.7){$\Lc$}
\put(6.9,-1.2){$\Cc$}
 \end{picture}
}
\label{symm1}
\ee
and
\be
 \mbox{
 \begin{picture}(200,70)(150,-10)
    \unitlength=0.9cm
  \thinlines
\put(4.2,1.5){$\Qc_{\kappa}$}
\put(6.85,1.5){$\Qc_{2\cdot\kappa}$}
\put(5.4,1.6){\vector(-1,0){0.4}}
\put(5.4,1.6){\vector(1,0){1.25}}
\put(5.7,1.8){$\Kc$}
\put(9.4,1.5){$\Wc(\D_{a,b})$}
\put(13.05,1.5){$\Wc(\D_{p-a,b})$}
\put(12.2,1.6){\vector(1,0){.65}}
\put(12.2,1.6){\vector(-1,0){1}}
\put(11.8,1.8){$\Mc$}
\put(4.2,0){$\Wc(\D_{a,b})$}
\put(7.9,0){$0$}
\put(6,0.1){\vector(1,0){1.65}}
\put(6.4,0.3){$\Kc,\Lc$}
\put(10,0.1){\vector(-1,0){1.65}}
\put(9,0.3){$\Mc$}
\put(10.2,0){$\R\in\ketw{\Jc^{\mathrm{irr}}_{p,p'}}\setminus\big\{\Wc(\D_{a,b})\big\}$}
 \end{picture}
}
\label{symm2}
\ee
depict the non-trivial actions of the maps
$\Kc,\Lc,\Mc,\Cc$ on $\ketw{\Jc^{\mathrm{conj}}_{p,p'}}$.
As indicated in Figure~\ref{CO}, the conjugation $\Cc$ does in general {\em not}
commute with the three maps $\Kc$, $\Lc$ and $\Mc$. Its inclusion in the
algebra in Figure~\ref{Cayley} would thus result in a {\em non-commutative}
composition algebra. Commutativity is respected
on $\ketw{\Jc^{\mathrm{out}}_{p,p'}}\cup\{\Wc(\D_{a,b})\}$, though.
In all instances, the fusion algebras to be discussed in the following {\em are} commutative.
\begin{figure}
$$
\renewcommand{\arraystretch}{1.5}
\begin{array}{c||cccccc}
  &\Qc_\kappa&\ketw{a,b}&\Wc(\D_{p-a,b})&\Wc(\D_{2p-a,b})&\Wc(\D_{3p-a,b})&\ketwast{a,b}
\\[4pt]
\hline \hline
\rule{0pt}{14pt}
 \Cc\circ\Kc
  &\Qc_{2\cdot\kappa}&\Wc(\D_{3p-a,b})&0&\Wc(\D_{3p-a,b})&\ketw{a,b}&\Wc(\D_{3p-a,b})
\\[4pt]
 \Kc\circ\Cc
  &\Qc_{2\cdot\kappa}&\Wc(\D_{3p-a,b})&0&\Wc(\D_{3p-a,b})&\ketwast{a,b}&\Wc(\D_{3p-a,b})
\\[4pt]
 \Cc\circ\Lc
  &\Qc_{\kappa}&\Wc(\D_{2p-a,b})&0&\ketw{a,b}&\Wc(\D_{3p-a,b})&\ketw{a,b}
\\[4pt]
 \Lc\circ\Cc
  &\Qc_{\kappa}&\ketwast{a,b}&0&\ketwast{a,b}&\Wc(\D_{3p-a,b})&\Wc(\D_{2p-a,b})
\\[4pt]
 \Cc\circ\Mc
  &0&\Wc(\D_{p-a,b})&\Wc(\D_{a,b})&0&0&0
\\[4pt]
 \Mc\circ\Cc
  &0&0&\Wc(\D_{a,b})&0&0&\Wc(\D_{p-a,b})
\end{array}
$$
\caption{Table indicating the results of acting with $\Cc\circ\Oc$ and 
$\Oc\circ\Cc$, for $\Oc\in\{\Kc,\Lc,\Mc\}$, on the various types of modules.}
\label{CO}
\end{figure}

\subsection{Symmetries and fusion rules}
\label{SecSymm}

The fusion algebra $\mathrm{Conj}[{\cal WLM}(p,p')]$
to be discussed presently is constructed as an extension
of the fundamental fusion algebra $\mathrm{Fund}[{\cal WLM}(p,p')]$ whose fusion rules 
\be
 \R_i\otimes\R_j=\bigoplus_k{\Nc_{i,j}}^k\R_k,\qquad\quad \R_i,\R_j,\R_k\in\ketw{\Jc^{\mathrm{fund}}_{p,p'}},\quad {\Nc_{i,j}}^k\in\mathbb{N}_0
\label{RRR}
\ee
are given in~\cite{Ras0812} and recalled in Appendix~\ref{AppFus}.
By an extension of a fusion algebra $\Ac$, we simply mean a fusion algebra containing $\Ac$ as
a non-trivial subalgebra.
\\[.2cm]
{\bf Proposition 1}\ \ \ Introducing 
$\mathrm{Conj}[{\cal WLM}(p,p')]=\big\langle\ketw{\Jc^{\mathrm{conj}}_{p,p'}}\big\rangle$ as an
extension of the fundamental fusion algebra (\ref{RRR}), by
requiring it to be $\Kc$-, $\Lc$- and $\Mc$-symmetric,
yields a unique fusion algebra. It is commutative and associative and has only
non-negative integer fusion multiplicities. The module $\ketw{1,1}$ is the unit.
\\[.2cm]
{\bf Proof}\ \ \ 
The uniqueness is an immediate consequence of the structure of the diagram
(\ref{symm1}) where every module in 
$\ketw{\Jc^{\mathrm{conj}}_{p,p'}}\setminus\ketw{\Jc^{\mathrm{fund}}_{p,p'}}$
can be written as $\Oc[\ketw{a,b}]$ for some $\Oc\in\Bc$ and some 
$\ketw{a,b}\in\ketw{\Jc^{\mathrm{fund}}_{p,p'}}$.
With $\Oc,\Oc',\Oc''\in\Bc$, associativity follows from
\bea
 &&\!\!\!\!
   \Big(\Oc\big[\ketw{a,b}\big]\otimes\Oc'\big[\ketw{a',b'}\big]\Big)\otimes\Oc''\big[\ketw{a'',b''}\big]
      =\Oc''\big[\Oc'\circ\Oc\big[\ketw{a,b}\otimes\ketw{a',b'}\big]\otimes\ketw{a'',b''}\big]\nn
  &&\qquad\qquad\qquad\qquad\qquad
    =\Oc''\circ\Oc'\circ\Oc\big[\ketw{a,b}\otimes\ketw{a',b'}\otimes\ketw{a'',b''}\big]\nn
  &&\qquad\qquad\qquad\qquad\qquad
    =\Oc\big[\ketw{a,b}\big]\otimes\Big(\Oc'\big[\ketw{a',b'}\big]\otimes\Oc''\big[\ketw{a'',b''}\big]\Big)
\eea
Commutativity follows similarly.
The fusion multiplicities are all taken from the set of fusion multiplicities
appearing in (\ref{RRR}) and are therefore non-negative integers.
Since every module $\R\in\ketw{\Jc^{\mathrm{conj}}_{p,p'}}$ can be written as
$\R=\Oc[\Fc]$ for some $\Oc\in\Bc\cup\{I\}$ and $\Fc\in\ketw{\Jc^{\mathrm{fund}}_{p,p'}}$,
the unital property of $\ketw{1,1}$ follows from
\be
 \ketw{1,1}\otimes\R=\ketw{1,1}\otimes\Oc[\Fc]=\Oc\big[\ketw{1,1}\otimes\Fc\big]
  =\Oc\big[\Fc\big]=\R
\ee
$\Box$
\\[.2cm]
As an immediate consequence of Proposition~1, we see that
\be
 \Oc\big[\R\big]\otimes\Oc'\big[\R'\big]=\Oc'\big[\R\big]\otimes\Oc\big[\R'\big],\qquad\quad
  \Oc,\Oc'\in\Bc,\quad \R,\R'\in\ketw{\Jc^{\mathrm{conj}}_{p,p'}}
\ee
Combined with the composition algebra of Figure~\ref{Cayley},
it also follows that, in addition to (\ref{RRR}), 
the fusion rules underlying $\mathrm{Conj}[{\cal WLM}(p,p')]$ are given 
by\footnote{Similar fusion rules have been conjectured independently by 
Simon Wood~\cite{Wood0907}.}
\bea
 && \Wc(\D_{2p-a,b})\otimes\Qc_\kappa=\ketw{a,b}\otimes\Qc_\kappa,\qquad
   \Wc(\D_{3p-a,b})\otimes\Qc_\kappa=\ketw{a,b}\otimes\Qc_{2\cdot\kappa}\nn
 &&\Wc(\D_{p-a,b})\otimes\Qc_\kappa=0,\qquad
  \ketwast{a,b}\otimes\Qc_\kappa=\ketw{a,b}\otimes\Qc_\kappa,\qquad\quad
    \Qc_\kappa\in\ketw{\Jc^{\mathrm{out}}_{p,p'}}
\label{fusconj1}
\eea
and (here written in `reverse order')
\bea
 0&=&\Wc(\D_{p-a,b})\otimes\Wc(\D_{2p-a',b'})=\Wc(\D_{p-a,b})\otimes\Wc(\D_{3p-a',b'})\nn
   &=&\Wc(\D_{p-a,b})\otimes\ketwast{a',b'}\nn
 \Kc\big[\ketw{a,b}\otimes\ketw{a',b'}\big]
 &=&\Wc(\D_{3p-a,b})\otimes\ketw{a',b'} =\Wc(\D_{2p-a,b})\otimes\Wc(\D_{3p-a',b'})\nn
 &=&\Wc(\D_{3p-a,b})\otimes\ketwast{a',b'}\nn
 \Lc\big[\ketw{a,b}\otimes\ketw{a',b'}\big]
 &=&\Wc(\D_{2p-a,b})\otimes\ketw{a',b'}\nn 
 \Lc^2\big[\ketw{a,b}\otimes\ketw{a',b'}\big]
 &=&\Wc(\D_{2p-a,b})\otimes\Wc(\D_{2p-a',b'})=\Wc(\D_{3p-a,b})\otimes\Wc(\D_{3p-a',b'})\nn
  &=&\Wc(\D_{2p-a,b})\otimes\ketwast{a',b'}
    =\ketwast{a,b}\otimes\ketw{a',b'} =\ketwast{a,b}\otimes\ketwast{a',b'}\nn
 \Mc\big[\ketw{a,b}\otimes\ketw{a',b'}\big]
 &=&\Wc(\D_{p-a,b})\otimes\ketw{a',b'}\nn
 \Mc^2\big[\ketw{a,b}\otimes\ketw{a',b'}\big]
 &=&\Wc(\D_{p-a,b})\otimes\Wc(\D_{p-a',b'})
\label{fusconj2}
\eea
These applications of $\Kc,\Lc,\Lc^2,\Mc,\Mc^2$ are trivially evaluated and their results are listed 
in Appendix~\ref{AppFusB} for completeness.
In particular, we find that
\bea
 \Wc(\D_{a,b})\otimes\Wc(\D_{a',b'})&=&\Mc^2\big[\ketw{p-a,b}\otimes\ketw{p-a',b'}\big]\nn
 &=&\Mc^2\big[\bigoplus_{a''=|a-a'|+1,\ \!{\rm by}\ \!2}^{p-|p-a-a'|-1}\
     \bigoplus_{b''=|b-b'|+1,\ \!{\rm by}\ \!2}^{p'-|p'-b-b'|-1}\ketw{a'',b''}\oplus\Qc\big]\nn
  &=&\bigoplus_{a''=|a-a'|+1,\ \!\mathrm{by}\ \!2}^{p-|p-a-a'|-1}\
   \bigoplus_{b''=|b-b'|+1,\ \!\mathrm{by}\ \!2}^{p'-|p'-b-b'|-1}\Wc(\D_{a'',b''})
\label{DabR}
\eea
where $\Qc\in\ketw{\Jc^{\mathrm{out}}_{p,p'}}$ is given in (\ref{sup0}). The irreducible
modules $\Wc(\D_{a,b})$ are thus seen to generate a fusion subalgebra isomorphic
to the fusion algebra of the usual {\em rational} minimal model 
${\cal M}(p,p')$~\cite{BPZ84,DiFMS}. This was also observed in~\cite{FGST0606}.

Proposition~1 implies the existence of a triplet of
symmetry-generating modules whose fusion rules correspond to the action of 
the three maps $\Kc$, $\Lc$ and $\Mc$.
\\[.2cm]
{\bf Proposition 2}\ \ \ For $p>1$, the $\Kc$-, $\Lc$- and $\Mc$-symmetries of
$\mathrm{Conj}[{\cal WLM}(p,p')]$ are governed by the
three modules $\Wc(\D_{3p-1,1})$, $\Wc(\D_{2p-1,1})$ and $\Wc(\D_{p-1,1})$, respectively,
in the sense that
\be
 \Kc\big[\R\big]=\Wc(\D_{3p-1,1})\otimes\R,\qquad
 \Lc\big[\R\big]=\Wc(\D_{2p-1,1})\otimes\R,\qquad
 \Mc\big[\R\big]=\Wc(\D_{p-1,1})\otimes\R
\label{K3L2M1}
\ee 
for all $\R\in\ketw{\Jc^{\mathrm{conj}}_{p,p'}}$.
\\[.2cm]
{\bf Proof}\ \ \ This is an immediate consequence of
\be
 \Oc\big[\R\big]=\Oc\big[\ketw{1,1}\otimes\R\big]=\Oc\big[\ketw{1,1}\big]\otimes\R,
  \qquad\quad\Oc\in\big\{\Kc,\Lc,\Mc\big\},\quad\R\in\ketw{\Jc^{\mathrm{conj}}_{p,p'}}
\ee
where
\be
 \Kc\big[\ketw{1,1}\big]=\Wc(\D_{3p-1,1}),\qquad
 \Lc\big[\ketw{1,1}\big]=\Wc(\D_{2p-1,1}),\qquad
 \Mc\big[\ketw{1,1}\big]=\Wc(\D_{p-1,1})
\label{KLM11}
\ee 
$\Box$
\\[.2cm]
For simple reference, we note that
\be
 \D_{kp-1,1}=\D_{1,kp'-1}=\frac{1}{4}(kp-2)(kp'-2)
\ee
In $\mathrm{Fund}[{\cal WLM}(p,p')]$, the two modules $\ketw{2,1}$ and $\ketw{1,2}$ 
(in addition to $\ketw{1,1}$) are naturally considered {\em fundamental}, cf. (\ref{Fund2112}).
In $\mathrm{Conj}[{\cal WLM}(p,p')]$ for $p>1$, on the other hand, we see that
the three irreducible modules $\Wc(\D_{kp-1,1})$, $k\in\mathbb{Z}_{1,3}$, too, should be regarded
as fundamental since
\be
 \mathrm{Conj}[{\cal WLM}(p,p')]=\big\langle\ketw{1,1},\ketw{2,1},\ketw{1,2},
  \Wc(\D_{p-1,1}),\Wc(\D_{2p-1,1}),\Wc(\D_{3p-1,1})\big\rangle
\label{Conj6}
\ee
To avoid confusion, we propose to refer to the six modules appearing explicitly
in (\ref{Conj6}) as {\em basic} modules. The fusion rules for these modules are
summarized in Appendix~\ref{AppBasic}.

Even though the conjugation $\Cc$ is an involution on $\ketw{\Jc^{\mathrm{conj}}_{p,p'}}$, it 
is {\em not} a symmetry of the fusion algebra $\mathrm{Conj}[{\cal WLM}(p,p')]$ for $p>1$.
This is illustrated by 
\be
 \Cc\big[\ketw{1,1}\big]\otimes\ketwast{1,1}=\ketwast{1,1}\neq\ketw{1,1}
  =\Cc\big[\ketw{1,1}\otimes\ketwast{1,1}\big]
\ee
Instead, one verifies the following weaker result.
\\[.2cm]
{\bf Proposition 3}\ \ \ 
\be
 \Fc\otimes\R^{\raisebox{0.25ex}{$\ast$}}=(\Fc\otimes\R)^{\raisebox{0.25ex}{$\ast$}},
   \qquad\quad \Fc\in\ketw{\Jc^{\mathrm{fund}}_{p,p'}},\quad
    \R\in\ketw{\Jc^{\mathrm{conj}}_{p,p'}}
\label{FRFR}
\ee

There are several results of the form appearing in (\ref{FRFR}), such as
$\Wc(\D_{3p-a,b})\otimes\Cc[\ketwast{a',b'}]=\Cc[\Wc(\D_{3p-a,b})\otimes\ketw{a',b'}]$, for example,
but we do not exhaust here the various possible extensions of Proposition~3.
We also note that $\R^{\raisebox{0.25ex}{$\ast$}}\otimes\R'^{\raisebox{0.25ex}{$\ast$}}$
is not, in general, equal to $(\R\otimes\R')^{\raisebox{0.25ex}{$\ast$}}$ as illustrated by
setting $\R=\ketw{1,1}$ and $\R'=\ketwast{1,1}$ for $p>1$
\be
 (\ketw{1,1})^{\raisebox{0.25ex}{$\ast$}}\otimes(\ketwast{1,1})^{\raisebox{0.25ex}{$\ast$}}
   =\ketwast{1,1}\neq\ketw{1,1}=(\ketw{1,1}\otimes\ketwast{1,1})^{\raisebox{0.25ex}{$\ast$}}
\ee

\subsection{Fusion algebra generated from irreducible modules}

A simple inspection of the fusion algebra $\mathrm{Conj}[{\cal WLM}(p,p')]$ reveals that
the modules of the form $\ketw{a,b}$ do {\em not} appear as the result of fusions involving
a module in $\ketw{\Jc^{\mathrm{conj}}_{p,p'}}\setminus\{\ketw{a,b}\}$.
One also observes that all but the modules $\ketw{a,b}$ are generated by repeated fusions
of the irreducible modules (\ref{irr}). This implies, in particular, 
that the fusion algebra $\big\langle\ketw{\Jc^{\mathrm{irr}}_{p,p'}}\big\rangle$
generated from repeated fusions of the irreducible modules is a {\em subalgebra}
of $\mathrm{Conj}[{\cal WLM}(p,p')]$. For $p>1$, this subalgebra does {\em not}
have a unit since $\ketw{1,1}$ is in the omitted set $\{\ketw{a,b}\}$. 

As already indicated, the spectrum of modules underlying the fusion algebra 
$\big\langle\ketw{\Jc^{\mathrm{irr}}_{p,p'}}\big\rangle$ is given by
$\ketw{\Jc^{\mathrm{conj}}_{p,p'}}\setminus\{\ketw{a,b}\}$ and is obviously
not invariant under conjugation for $p>1$. The minimal conjugation-invariant
extension of this set is $\ketw{\Jc^{\mathrm{conj}}_{p,p'}}$. Thus, requiring that
the spectrum is invariant under conjugation brings back the fusion algebra
$\mathrm{Conj}[{\cal WLM}(p,p')]$ and hence the identity $\ketw{1,1}$.
We recall, though, that conjugation is {\em not} a symmetry of
this fusion algebra for $p>1$, cf. the discussion leading up to Proposition~3.

\subsection{On fusion subalgebras and quotients}
\label{SecProp}

So far, we have encountered
\be
 \big\langle\ketw{\Jc^{\mathrm{out}}_{p,p'}}\big\rangle
   \subset\big\langle\ketw{\Jc^{\mathrm{fund}}_{p,p'}}\big\rangle
   \subset\big\langle\ketw{\Jc^{\mathrm{conj}}_{p,p'}}\big\rangle,\qquad
 \big\langle\ketw{\Jc^{\mathrm{out}}_{p,p'}}\big\rangle
   \subset\big\langle\ketw{\Jc^{\mathrm{irr}}_{p,p'}}\big\rangle
   \subset\big\langle\ketw{\Jc^{\mathrm{conj}}_{p,p'}}\big\rangle
\ee
as sequences of fusion (sub)algebras.
For $p>1$, the two sets $\ketw{\Jc^{\mathrm{fund}}_{p,p'}}$ and 
$\ketw{\Jc^{\mathrm{irr}}_{p,p'}}$ are not related by $\subset$, while
the various extensions of $\ketw{\Jc^{\mathrm{out}}_{p,p'}}$ are trivial for $p=1$
\be
 \big\langle\ketw{\Jc^{\mathrm{out}}_{1,p'}}\big\rangle
   =\big\langle\ketw{\Jc^{\mathrm{fund}}_{1,p'}}\big\rangle
   =\big\langle\ketw{\Jc^{\mathrm{irr}}_{1,p'}}\big\rangle
   =\big\langle\ketw{\Jc^{\mathrm{conj}}_{1,p'}}\big\rangle
\label{p1}
\ee
The fusion algebra $\mathrm{Conj}[{\cal WLM}(p,p')]$
contains many other fusion subalgebras than the ones listed above.
As some of these are ideals, one may
also consider the corresponding quotient structures.
Here, we address some of these subalgebras and quotients.

First, the fusion algebra generated by the irreducible modules $\Wc(\D_{a,b})$
is such an ideal. The quotient
$\big\langle\ketw{\Jc^{\mathrm{conj}}_{p,p'}}\big\rangle/(\{\Wc(\D_{a,b})\})$
is equivalent to the fusion algebra constructed as in Proposition~1 if one works
with $\ketw{\Jc^{\mathrm{conj}}_{p,p'}}\setminus\{\Wc(\D_{a,b})\}$ and refrains
from imposing the $\Mc$-symmetry. As a non-trivial subalgebra, it contains
the fusion algebra 
$\big\langle\ketw{\Jc^{\mathrm{out}}_{p,p'}}
\cup\{\ketwast{a,b},\Wc(\D_{(\kappa+1)p-a})\}\big\rangle$, which, for $p>1$,
does not have a unit.
Likewise, refraining from imposing the $\Kc$- or $\Lc$-symmetry, or any combination of the
three symmetries, on the corresponding subset of $\ketw{\Jc^{\mathrm{conj}}_{p,p'}}$, 
yields a fusion subalgebra of $\mathrm{Conj}[{\cal WLM}(p,p')]$.
Such a subalgebra can, in general, not be described as a quotient of 
$\mathrm{Conj}[{\cal WLM}(p,p')]$ simply because the
omitted modules do not generate an ideal. For example, let us consider
the situation arising when leaving out the $\Lc$-symmetry
and omitting the modules $\Wc(\D_{2p-a,b})$. This yields a perfectly
well-defined fusion subalgebra even though $\Lc[\Wc(\D_{2p-a,b})]=\ketwast{a,b}$
prevents the modules $\Wc(\D_{2p-a,b})$ from generating an ideal
of $\mathrm{Conj}[{\cal WLM}(p,p')]$.

\section{Conjecture and comparison with known results}

Denoting by $\mathrm{Irr}[{\cal WLM}(p,p')]$ the fusion algebra generated by 
repeated fusions of the irreducible modules according to the 
Nahm-Gaberdiel-Kausch algorithm~\cite{Nahm9402,GK9604},
we conjecture that it is identical to the fusion algebra 
$\big\langle\ketw{\Jc^{\mathrm{irr}}_{p,p'}}\big\rangle$ discussed above.
\\[.2cm]
{\bf Conjecture}\ \ \ 
\be
 \mathrm{Irr}[{\cal WLM}(p,p')]=\big\langle\ketw{\Jc^{\mathrm{irr}}_{p,p'}}\big\rangle
\label{conj}
\ee

In support of this assertion, we first note that the irreducible module
$\Wc(\D_{a,b})$ of the ${\cal W}$-extended Virasoro algebra is, in fact, an irreducible
module of the Virasoro algebra itself, that is, 
\be
 \Wc(\D_{a,b})=\Vc(\D_{a,b})
\ee
{}From the Nahm-Gaberdiel-Kausch algorithm, one thus recovers 
the usual rational minimal-model Virasoro fusion rules (\ref{DabR}).

Due to (\ref{p1}), the comparison of our proposal for $p=1$ with the 
results~\cite{GK9606,GR0707}
obtained by application of the Nahm-Gaberdiel-Kausch algorithm was already performed 
in~\cite{PRR0803}.
For $p>1$, the situation is considerably more complicated and much less is known
about the implications of the algorithm.
Following the earlier work~\cite{EF0604} on fusion of irreducible Virasoro modules,
the fusion algebra generated by repeated fusions of the 13 irreducible modules
appearing in ${\cal WLM}(2,3)$ was recently worked out in~\cite{GRW0905}.
The focus there was on a conjugation-invariant spectrum, and we
have verified that $\mathrm{Conj}[{\cal WLM}(2,3)]$ indeed corresponds to their results.
For ease of comparison, we note that
\be
 \Wc=\ketw{1,1},\qquad \Qc=\ketw{1,2},\qquad 
  \Wc^{\raisebox{0.25ex}{$\ast$}}_{\phantom{\Wc}}=\ketwast{1,1},
   \qquad \Qc^{\raisebox{0.25ex}{$\ast$}}_{\phantom{\Wc}}=\ketwast{1,2}
\label{WQ}
\ee
in the notation of~\cite{GRW0905}. 
This, of course, presupposes that the notion of the {\em conjugate} of a module
in $\ketw{\Jc^{\mathrm{conj}}_{p,p'}}$ is the same as the one employed in~\cite{GRW0905}.
This is easily verified. Essential aspects of the role played by $\Wc(\D_{3p-1,1})=\Wc(7)$ 
in ${\cal WLM}(2,3)$ were described in~\cite{GRW0905}. 
In particular, it was found that the modules in
$\ketw{\Jc^{\mathrm{out}}_{2,3}}
\cup\{\Wc(5),\Wc(7)\}\cup\{\Wc^{\raisebox{0.25ex}{$\ast$}},\Qc^{\raisebox{0.25ex}{$\ast$}}\}$
are organized in pairs with respect to fusion with $\Wc(7)$.
From our perspective, this corresponds to the maps
\be
 \mbox{
 \begin{picture}(290,16)(0,0)
    \unitlength=0.9cm
  \thinlines
\put(0.2,0){$\Qc_{\kappa}$}
\put(2.85,0){$\Qc_{2\cdot\kappa}$}
\put(1.4,0.1){\vector(-1,0){0.4}}
\put(1.4,0.1){\vector(1,0){1.25}}
\put(1.7,0.3){$\Kc$}
\put(5.9,0){$\Wc(\D_{3p-a,b})$}
\put(10.3,0){$\ketwast{a,b}$}
\put(9.5,0.1){\vector(-1,0){1.15}}
\put(9.5,0.1){\vector(1,0){.5}}
\put(9.05,.3){$\Kc$}
 \end{picture}
}
\ee 
where $\Qc_\kappa\in\ketw{\Jc^{\mathrm{out}}_{2,3}}$, $a\in\mathbb{Z}_{1,p-1}=\{1\}$
and $b\in\mathbb{Z}_{1,p'-1}=\{1,2\}$.

Adopting some further terminology used in~\cite{GRW0905}, 
though without going into details, we find that the set of modules which have 
a {\em dual} module is given by $\ketw{\Jc^{\mathrm{fund}}_{p,p'}}$ and recall that they
generate a closed fusion algebra, namely $\mathrm{Fund}[{\cal WLM}(p,p')]$.
Believing that the modules in $\ketw{\Jc^{\mathrm{fund}}_{p,p'}}$ are, in fact, self-dual,
it follows that the set of self-conjugate and self-dual modules is given by
$\ketw{\Jc^{\mathrm{out}}_{p,p'}}$. This is exactly the set of modules
naturally associated with ${\cal W}$-invariant boundary conditions. 
They, too, generate a closed fusion algebra, namely $\mathrm{Out}[{\cal WLM}(p,p')]$.

\section{Polynomial fusion rings}

The fusion algebra
\be
 \phi_i\otimes\phi_j\ =\ \bigoplus_{k\in\mathcal{J}}{N_{i,j}}^k\phi_k,\hspace{1cm}i,j\in\mathcal{J}
\ee
of a {\em rational} conformal field theory is finite and
can be represented by a commutative matrix algebra $\langle N_i;\ i\in\mathcal{J}\rangle$ 
where the entries of the $|\mathcal{J}|\times|\mathcal{J}|$ matrix $N_i$ are
\be
 {(N_i)_j}^k\ =\ {N_{i,j}}^k,\hspace{1cm}i,j,k\in\mathcal{J}
\ee
and where the fusion multiplication $\otimes$ has been replaced by ordinary matrix multiplication.
In~\cite{Gep91}, Gepner found that every such algebra is isomorphic to a ring
of polynomials in a finite set of variables modulo an ideal defined as the vanishing conditions
of a finite set of polynomials in these variables. He also conjectured that this ideal of constraints 
corresponds to the local extrema of a potential, see~\cite{DiFZ9211,Aha9301,BR0602} for
further elaborations on this conjecture.

We extended Gepner's result to the fundamental fusion algebra
$\mathrm{Fund}[{\cal WLM}(p,p')]$ in~\cite{Ras0812} where we found that
\be
 \mathrm{Fund}[{\cal WLM}(p,p')]
   \simeq\mathbb{C}[X,Y]\big/\big(P_{p}(X),P_{p'}(Y),P_{p,p'}(X,Y)\big)
\label{FundPol}
\ee
Here,
\be
 P_n(x)=2\Big(T_{2n}\big(\frac{x}{2}\big)-1\Big)U_{n-1}\big(\frac{x}{2}\big),\qquad\quad
 P_{n,n'}(x,y)=\Big(T_n\big(\frac{x}{2}\big)-T_{n'}\big(\frac{y}{2}\big)\Big)
   U_{n-1}\big(\frac{x}{2}\big)U_{n'-1}\big(\frac{y}{2}\big)
\label{Pn}
\ee
where $T_n(x)$ and $U_n(x)$ are Chebyshev polynomials of the first and second kind,
respectively.
The isomorphism in (\ref{FundPol}) reads 
\bea
 \ketw{a,b}&\leftrightarrow&
      U_{a-1}\big(\frac{X}{2}\big)U_{b-1}\big(\frac{Y}{2}\big)\nn
 \Wc(\D_{\kappa p,s})&\leftrightarrow&
      \frac{1}{\kappa}U_{\kappa p-1}\big(\frac{X}{2}\big)U_{s-1}\big(\frac{Y}{2}\big)\nn
 \Wc(\D_{a,\kappa p'})&\leftrightarrow&
      \frac{1}{\kappa}U_{a-1}\big(\frac{X}{2}\big)U_{\kappa p'-1}\big(\frac{Y}{2}\big)\nn
 \ketw{\R_{\kappa p,s}^{a,0}}&\leftrightarrow&
     \frac{2}{\kappa}T_a\big(\frac{X}{2}\big)U_{\kappa p-1}\big(\frac{X}{2}\big)
        U_{s-1}\big(\frac{Y}{2}\big)\nn
 \ketw{\R_{r,\kappa p'}^{0,b}}&\leftrightarrow&
     \frac{2}{\kappa}U_{r-1}\big(\frac{X}{2}\big) 
       T_b\big(\frac{Y}{2}\big)U_{\kappa p'-1}\big(\frac{Y}{2}\big)\nn
 \ketw{\R_{\kappa p,p'}^{a,b}}&\leftrightarrow&
    \frac{4}{\kappa}T_a\big(\frac{X}{2}\big)
    U_{\kappa p-1}\big(\frac{X}{2}\big)
    T_b\big(\frac{Y}{2}\big)U_{p'-1}\big(\frac{Y}{2}\big)
\label{abR4}
\eea
where it is noted that
\be
 U_{\kappa p-1}\big(\frac{X}{2}\big)U_{p'-1}\big(\frac{Y}{2}\big)
  \equiv U_{p-1}\big(\frac{X}{2}\big)U_{\kappa p'-1}\big(\frac{Y}{2}\big)\qquad
    (\mathrm{mod}\ P_{p,p'}(X,Y))
\ee
for example.

We now wish to show that the fusion algebra $\mathrm{Conj}[{\cal WLM}(p,p')]$
also admits a polynomial-ring description.
For $p=1$, this is trivially true since 
$\mathrm{Conj}[{\cal WLM}(1,p')]=\mathrm{Fund}[{\cal WLM}(1,p')]$.
In the following, we will therefore assume that $p>1$.

According to Proposition~1, the fusion algebra 
$\mathrm{Conj}[{\cal WLM}(p,p')]$ is constructed as an extension
of $\mathrm{Fund}[{\cal WLM}(p,p')]$ where the extension, according to
Proposition~2, is governed by the three irreducible modules 
$\Wc(\D_{kp-1,1})$, $k\in\mathbb{Z}_{1,3}$. 
Following from the diagrams (\ref{symm1}), (\ref{symm2}) and the ensuing composition algebra
in Figure~\ref{Cayley}, we find that $\mathrm{Conj}[{\cal WLM}(p,p')]$ can
be described by a quotient polynomial ring with five generators, cf. (\ref{Conj6}).
\\[.2cm]
{\bf Proposition 4}\ \ \ For $p>1$, the fusion algebra $\mathrm{Conj}[{\cal WLM}(p,p')]$
is isomorphic to the quotient polynomial ring
\be
 \mathrm{Conj}[{\cal WLM}(p,p')]\simeq\mathbb{C}[X,Y,K,L,M]/\Ic_{p,p'}
\label{ConjPol}
\ee
where $\Ic_{p,p'}$ is the ideal defined by the vanishing conditions
\bea
  0
    &=&\Big(K-T_p\big(\frac{X}{2}\big)\Big)U_{p-1}\big(\frac{X}{2}\big)
      =\Big(K-T_{p'}\big(\frac{Y}{2}\big)\Big)U_{p'-1}\big(\frac{Y}{2}\big)
       =(L-1)U_{p-1}\big(\frac{X}{2}\big)=(L-1)U_{p'-1}\big(\frac{Y}{2}\big)\nn
    &=&M^2-MU_{p-2}\big(\frac{X}{2}\big)=M^2-MU_{p'-2}\big(\frac{Y}{2}\big)
      =K(L-1)=K^2-L^2=KM=LM
\label{Ipp}
\eea
The isomorphism in (\ref{ConjPol}) is given by (\ref{abR4}) supplemented by
\bea
  \Wc(\D_{p-a,b})&\leftrightarrow&MU_{a-1}\big(\frac{X}{2}\big)U_{b-1}\big(\frac{Y}{2}\big)\nn
  \Wc(\D_{2p-a,b})&\leftrightarrow&LU_{a-1}\big(\frac{X}{2}\big)U_{b-1}\big(\frac{Y}{2}\big)\nn
  \Wc(\D_{3p-a,b})&\leftrightarrow&KU_{a-1}\big(\frac{X}{2}\big)U_{b-1}\big(\frac{Y}{2}\big)\nn
  \ketwast{a,b}&\leftrightarrow&L^2U_{a-1}\big(\frac{X}{2}\big)U_{b-1}\big(\frac{Y}{2}\big)
\label{ConjIso}
\eea
{\bf Proof}\ \ \ 
With 
\be
 K\leftrightarrow\Wc(\D_{3p-1,1}),\qquad
 L\leftrightarrow\Wc(\D_{2p-1,1}),\qquad
 M\leftrightarrow\Wc(\D_{p-1,1})
\label{KLMWD}
\ee
it follows that $\mathrm{Conj}[{\cal WLM}(p,p')]$ is isomorphic to the quotient polynomial
ring in $X$, $Y$, $K$, $L$ and $M$ whose defining ideal can be described
by supplementing the conditions in (\ref{FundPol}) with
the conditions following from translating the various arrows (including the trivial identity maps)
in the diagrams (\ref{symm1}) and (\ref{symm2}) into polynomial constraints.
Completing the proof thus amounts to verifying that the set of conditions
in (\ref{Ipp}) is {\em necessary} and
{\em sufficient} to characterize this ideal. This is straightforwardly done.
Here, we only include a couple of these verifications as the remaining ones are treated
similarly. First, that a condition is necessary means that it is a consequence of the
conditions given in (\ref{FundPol}) combined with the ones following from the 
translation procedure. From $\Kc\big[\Wc(\D_{p,1})\big]=\Wc(\D_{2p,1})$,
for example, we thus conclude that 
$KU_{p-1}\big(\frac{X}{2}\big)=\frac{1}{2}U_{2p-1}\big(\frac{X}{2}\big)$.
Using the identity $U_{2p-1}(x)=2T_p(x)U_{p-1}(x)$,
we immediately recognize the first condition appearing in (\ref{Ipp}).
To illustrate that the conditions in (\ref{Ipp}) are sufficient, we observe that 
\be
 MU_{\kappa p-1}\big(\frac{X}{2}\big)\equiv0\qquad 
   (\mathrm{mod}\ (L-1)U_{p-1}\big(\frac{X}{2}\big),LM)
\ee
where the congruence for $\kappa=2$ is a simple consequence of the one for $\kappa=1$.
Multiplied by $U_{s-1}\big(\frac{Y}{2}\big)$, this corresponds to 
$\Mc\big[\Wc(\D_{\kappa p,s})\big]=0$.
Using the identity
\be
 U_{p-2}(x)U_{a-1}(x)=U_{p-a-1}(x)
  +\sum_{n=-(a-2),\ \!{\rm by}\ \!2}^{a-2}T_{|n|}(x)U_{p-1}(x)
\ee
we subsequently find that
\be
 M^2U_{a-1}\big(\frac{X}{2}\big)
   \equiv MU_{p-a-1}\big(\frac{X}{2}\big),\qquad
     (\mathrm{mod}\ M^2-MU_{p-2}\big(\frac{X}{2}\big),MU_{p-1}\big(\frac{X}{2}\big))
\label{M2U}
\ee
which, multiplied by $U_{b-1}\big(\frac{Y}{2}\big)$,
corresponds to $\Mc^2\big[\ketw{a,b}\big]=\Mc\big[\ketw{p-a,b}\big]$ 
which itself comes from $\Mc\big[\Wc(\D_{p-a,b})\big]=\Wc(\D_{a,b})$.
Let us also consider 
$\Kc\big[\ketw{\R_{\kappa p,p'}^{a,b}}\big]=\big[\ketw{\R_{(2\cdot\kappa)p,p'}^{a,b}}\big]$
corresponding to
\be
 K\frac{4}{\kappa}T_a\big(\frac{X}{2}\big)U_{\kappa p-1}\big(\frac{X}{2}\big)
   T_b\big(\frac{Y}{2}\big)U_{p'-1}\big(\frac{Y}{2}\big)\equiv
 \frac{4}{2\cdot\kappa}T_a\big(\frac{X}{2}\big)U_{(2\cdot\kappa)p-1}\big(\frac{X}{2}\big)
   T_b\big(\frac{Y}{2}\big)U_{p'-1}\big(\frac{Y}{2}\big)
\ee
For $\kappa=1$, this follows immediately from 
$0=\big(K-T_p\big(\frac{X}{2}\big)\big)U_{p-1}\big(\frac{X}{2}\big)$. For $\kappa=2$,
it follows from $P_p(X)=0$ which is not, though, in the set (\ref{Ipp}). 
However, we wish to emphasize that, not only $P_p(X)=0$, but all three conditions 
appearing in (\ref{FundPol}) are consequences of the conditions in (\ref{Ipp}). This follows from
\be
 P_{p,p'}(X,Y)=\Big(K-T_{p'}\big(\frac{Y}{2}\big)\Big)U_{p-1}\big(\frac{X}{2}\big)
     U_{p'-1}\big(\frac{Y}{2}\big)
   -\Big(K-T_p\big(\frac{X}{2}\big)\Big)U_{p-1}\big(\frac{X}{2}\big)U_{p'-1}\big(\frac{Y}{2}\big)
  \equiv0
\ee
where the congruence is modulo 
$\big(K-T_{p}\big(\frac{X}{2}\big)\big)U_{p-1}\big(\frac{X}{2}\big)$
and $\big(K-T_{p'}\big(\frac{Y}{2}\big)\big)U_{p'-1}\big(\frac{Y}{2}\big)$, and from
\be
 P_p(X)=4\Big(T_p^2\big(\frac{X}{2}\big)-1\Big)U_{p-1}\big(\frac{X}{2}\big)\equiv0
\ee
where the congruence is modulo $(L-1)U_{p-1}\big(\frac{X}{2}\big)$, $K^2-L^2$ and 
$\big(K-T_{p}\big(\frac{X}{2}\big)\big)U_{p-1}\big(\frac{X}{2}\big)$.
The condition for $P_{p'}(Y)$ follows similarly, of course.
\\
$\Box$
\\[.2cm]
Just as $P_p(X)$ and $P_{p'}(Y)$ are the minimal polynomials of $X$ and $Y$ modulo 
$\Ic_{p,p'}$, we see that $K(K^2-1)$, $L^2(L-1)$ and $M(M^2-1)$ 
are the minimal polynomials of $K$, $L$ and $M$, respectively. Indeed, 
using (\ref{M2U}), in particular, we have
\be 
 K^3\equiv KL^2\equiv KL\equiv K,\qquad
 L^3\equiv K^2L\equiv K^2\equiv L^2,\qquad
 M^3\equiv M^2U_{p-2}\big(\frac{X}{2}\big)\equiv M
\ee
modulo $\Ic_{p,p'}$.

{}From the analysis above, we extract
the conditions linking the modules $\Wc(\D_{a,b})$ to each other
\be
 0=MU_{p-1}\big(\frac{X}{2}\big)=MU_{p'-1}\big(\frac{Y}{2}\big)
   =M\Big(U_{p-2}\big(\frac{X}{2}\big)-U_{p'-2}\big(\frac{Y}{2}\big)\Big)
\label{MUMUMU}
\ee
Up to the factors of $M$, these are recognized as the standard
conditions defining the quotient polynomial ring associated with the {\em rational}
minimal models, see~\cite{DiFMS,Ras0812}, for example. 
This should not, though, come as a surprise since we have already
realized that the fusion subalgebra generated by the irreducible modules
$\Wc(\D_{a,b})$ satisfy the usual minimal-model fusion rules (\ref{DabR}).

Since the minimal fusion algebra generated from repeated fusions of the
irreducible modules,
$\big\langle\ketw{\Jc^{\mathrm{irr}}_{p,p'}}\big\rangle$, does not have a unit,
it cannot be isomorphic to a quotient polynomial ring.
However, since this fusion algebra is a subalgebra of 
$\mathrm{Conj}[{\cal WLM}(p,p')]=\big\langle\ketw{\Jc^{\mathrm{conj}}_{p,p'}}\big\rangle$,
it is isomorphic to a {\em subring} of the quotient polynomial ring
appearing in (\ref{ConjPol}). This subring is obtained by omitting the polynomials
$U_{a-1}\big(\frac{X}{2}\big)U_{b-1}\big(\frac{Y}{2}\big)$ {\em themselves}
from the ambient ring while keeping their products with other non-trivial polynomials.
We see that this corresponds to 
eliminating the identity map $\Ic$ from the allowed
operations on $\ketw{\Jc^{\mathrm{conj}}_{p,p'}}$ when constructing
the composition algebra in Figure~\ref{Cayley}. This elimination procedure 
is algebraically well-defined, cf. the closure of the composition algebra and
the discussion following (\ref{five}).
We also recall from~\cite{Ras0812} that omitting the polynomials 
$U_{a-1}\big(\frac{X}{2}\big)U_{b-1}\big(\frac{Y}{2}\big)$ from
the quotient polynomial ring in (\ref{FundPol}) yields a well-defined
subring isomorphic to the fusion algebra
$\mathrm{Out}[{\cal WLM}(p,p')]$ of the modules naturally associated with 
${\cal W}$-invariant boundary conditions.

\section{Concluding remarks}

Based on symmetry principles, we have derived a fusion algebra 
$\langle\ketw{\Jc_{p,p'}^{\mathrm{irr}}}\rangle$ generated from repeated fusions
of the irreducible modules appearing in the ${\cal W}$-extended logarithmic minimal model
${\cal WLM}(p,p')$. In addition to the irreducible 
modules themselves (\ref{irr}), closure of the fusion algebra requires 
the participation of a variety of reducible yet indecomposable modules. 
We conjecture that this fusion algebra is the same as the fusion algebra
$\mathrm{Irr}[{\cal WLM}(p,p')]$, also generated by repeated fusions of the irreducible modules,
but obtained by application of the Nahm-Gaberdiel-Kausch algorithm. In support of this
conjecture, we find that the two fusion algebras agree
for ${\cal WLM}(1,p')$~\cite{GK9606,GR0707,PRR0803} 
and for ${\cal WLM}(2,3)$~\cite{GRW0905}. 
For $p>1$, our fusion algebra does not contain an algebra unit.
Requiring that the spectrum of modules is invariant under a natural notion of conjugation, 
however, introduces an additional
$(p-1)(p'-1)$ reducible yet indecomposable rank-1 modules, among which
the identity is found. This bigger set of indecomposable modules is denoted by
$\ketw{\Jc_{p,p'}^{\mathrm{conj}}}$. The corresponding fusion
algebra $\mathrm{Conj}[{\cal WLM}(p,p')]$ is invariant under the symmetries $\Kc$, $\Lc$ and 
$\Mc$ and contains $\mathrm{Irr}[{\cal WLM}(p,p')]$ 
(or strictly speaking $\langle\ketw{\Jc_{p,p'}^{\mathrm{irr}}}\rangle$) as a subalgebra. 
These symmetry generators are maps from
$\ketw{\Jc_{p,p'}^{\mathrm{conj}}}$ to itself. Their actions on $\mathrm{Conj}[{\cal WLM}(p,p')]$ 
are shown to be generated by fusions with the three irreducible
modules of conformal weights $\D_{kp-1,1}$, $k=3,2,1$, respectively.
We have also identified a polynomial fusion ring isomorphic to the fusion 
algebra $\mathrm{Conj}[{\cal WLM}(p,p')]$. For $p>1$, it has five generators corresponding
to the two fundamental modules and the three symmetry generators.
The fusion algebra $\langle\ketw{\Jc_{p,p'}^{\mathrm{irr}}}\rangle$ is isomorphic to
a particular subring of the polynomial fusion ring.

The reader may wonder about our motivation for introducing the three maps $\Kc$, $\Lc$ and
$\Mc$ when the objective was to determine the fusion algebra generated by repeated fusions
of the irreducible modules. 
First, since $\Wc(\D_{a,b})=\Vc(\D_{a,b})$, we want the irreducible modules $\Wc(\D_{a,b})$ 
to generate a fusion subalgebra isomorphic to the fusion algebra
of the Virasoro minimal model ${\cal M}(p,p')$.
Second, from the success of $\mathrm{Fund}[{\cal WLM}(p,p')]$ as an ambient fusion algebra
hosting the fusion algebra $\mathrm{Out}[{\cal WLM}(p,p')]$ generated by the modules
associated with boundary conditions, we expect to encounter a fusion algebra generated 
by a small number of basic modules. As in the case of $\mathrm{Fund}[{\cal WLM}(p,p')]$, this 
fusion algebra may not be the sought-after fusion algebra itself ($\mathrm{Irr}[{\cal WLM}(p,p')]$), 
but rather an extension thereof ($\mathrm{Conj}[{\cal WLM}(p,p')]$). 
Third, examinations like (\ref{22}) reveal that the contragredient modules $\ketwast{a,b}$ are 
generated, and we are led to consider the set $\ketw{\Jc_{p,p'}^{\mathrm{conj}}}$ of 
indecomposable modules.
Fourth, we wish to preserve as much as possible the factorization enjoyed by the indecomposable 
modules of the fundamental fusion algebra where every module can be written as the
fusion of a ``horizontal" and a ``vertical" module: 
$\ketw{\R_{\rho,\sigma}^{\al,\beta}}
=\ketw{\R_{\rho,1}^{\al,0}}\otimes\ketw{\R_{1,\sigma}^{0,\beta}}$.
Supported by explicit evaluations, we then made the ansatz that the set of basic modules is 
given by the ones appearing explicitly in (\ref{Conj6}), that is,
\be
 \{\ketw{1,1},\ketw{2,1},\ketw{1,2},
  \Wc(\D_{p-1,1}),\Wc(\D_{2p-1,1}),\Wc(\D_{3p-1,1})\}
\ee
and that their fusion rules are the ones given in Figure~\ref{pgeq3} and \ref{peq2}.
For $p>2$, we thus have
\bea
 \ketw{1,2}\otimes\Wc(\D_{kp-1,1})\!&=&\!\Wc(\D_{kp-1,2})\nn
 \ketw{2,1}\otimes\Wc(\D_{1,kp'-1})\!&=&\!\ketw{2,1}\otimes\Wc(\D_{kp-1,1})
  \;=\;\Wc(\D_{kp-2,1})\;=\;\Wc(\D_{2,kp'-1})
\eea
for example, resembling the aforementioned factorization. 
Everything else is fixed by requiring associativity of the fusion algebra, the one called
$\mathrm{Conj}[{\cal WLM}(p,p')]$. The subalgebra generated from repeated fusions of the
irreducible modules is subsequently identified straightforwardly.
Since universality, as opposed to model-specific properties, is likely to be manifest when the basic 
rules of the game are expressed in terms of symmetry principles, we found it natural to try
to translate the fusion rules into such principles thereby introducing $\Kc$, $\Lc$ and $\Mc$. 
Once identified, these symmetry generators illuminate quite clearly the structure of the fusion 
algebra.

In summary, we have verified that our proposals provide well-defined fusion algebras 
($\langle\ketw{\Jc_{p,p'}^{\mathrm{irr}}}\rangle$ and $\mathrm{Conj}[{\cal WLM}(p,p')]$,
in particular) and that they reproduce all known results in this regard. 
We are not, though, making any claims of uniqueness of the constructions, but do conjecture that
$\langle\ketw{\Jc_{p,p'}^{\mathrm{irr}}}\rangle$ is identical to
$\mathrm{Irr}[{\cal WLM}(p,p')]$ obtained by application of the Nahm-Gaberdiel-Kausch algorithm. 
As generated from the minimal conjugation-invariant extension $\ketw{\Jc_{p,p'}^{\mathrm{conj}}}$
of the spectrum $\ketw{\Jc_{p,p'}^{\mathrm{irr}}}$, we furthermore conjecture that the $\Kc$-, 
$\Lc$- and $\Mc$-invariant fusion algebra $\mathrm{Conj}[{\cal WLM}(p,p')]$ is identical to the 
similar extension obtained by application of the Nahm-Gaberdiel-Kausch algorithm.

\subsection*{Note added}

After the present work appeared on the arXiv, the paper~\cite{Wood0907} appeared
on the arXiv. As a continuation of the work~\cite{GRW0905}, it also addresses the fusion
algebra generated from repeated fusions of irreducible modules, and it presents 
conjectured fusion rules similar to the ones proposed in Section~\ref{SecSymm}.

\subsection*{Acknowledgments}

This work is supported by the Australian Research Council. 
The author thanks Matthias Gaberdiel, Paul A. Pearce, David Ridout and
Simon Wood for helpful comments.

\appendix
\section{Fusion rules of $\mathrm{Conj}[{\cal WLM}(p,p')]$}

\subsection{Fundamental fusion algebra $\mathrm{Fund}[{\cal WLM}(p,p')]$}
\label{AppFus}

Here, we summarize the fusion rules, obtained in~\cite{Ras0805,Ras0812}, underlying the 
fusion algebra $\mathrm{Fund}[{\cal WLM}(p,p')]$ as given in (\ref{WfusFund}).
To this end, by a direct sum of representations $A_n$
with unspecified lower summation bound, we mean the direct sum in steps of 2
whose lower bound is given by the parity of the upper bound
\be
 \bigoplus_{n}^{N}A_n\ =\ \bigoplus_{n=\frac{1}{2}(1-(-1)^N),\ \!{\rm by}\ \!2}^{N}\!\! 
  A_n,\qquad\quad N\in\mathbb{Z}
\label{sum2}
\ee
This direct sum vanishes for negative $N$. 
For simplicity, and in compliance with the notation of~\cite{Ras0812},
we write $\ketw{\R_{\rho,\sigma}^{0,0}}=\ketw{\rho,\sigma}$,
$\ketw{\kappa p,s}=\Wc(\D_{\kappa p,s})$ and $\ketw{r,\kappa p'}=\Wc(\D_{r,\kappa p'})$.
Now, the fusions involving the module $\ketw{a,b}$ are given by
\bea
 \ketw{a,b}\otimes\ketw{a',b'}&=&
   \bigoplus_{i=|a-a'|+1,\ \!{\rm by}\ \!2}^{p-|p-a-a'|-1}
     \Big\{\bigoplus_{j=|b-b'|+1,\ \!{\rm by}\ \!2}^{p'-|p'-b-b'|-1}
  \!\!\!\ketw{i,j}\Big\}\oplus\!
  \bigoplus_{\al}^{a+a'-p-1}\!
   \Big\{\bigoplus_{j=|b-b'|+1,\ \!{\rm by}\ \!2}^{p'-|p'-b-b'|-1}\ketw{\R_{p,j}^{\al,0}}\Big\}\nn
 &&\oplus\!\bigoplus_{\beta}^{b+b'-p'-1}\!
   \Big\{\bigoplus_{i=|a-a'|+1,\ \!{\rm by}\ \!2}^{p-|p-a-a'|-1}\ketw{\R_{i,p'}^{0,\beta}}\Big\}
  \oplus\bigoplus_{\al}^{a+a'-p-1}
    \Big\{\bigoplus_{\beta}^{b+b'-p'-1}\!\ketw{\R_{p,p'}^{\al,\beta}}\Big\}
\label{sup0}
\eea
and
\bea
 \ketw{a,b}\otimes\ketw{\kappa p,b'}&=&\bigoplus_{\al}^{a-1}\Big\{
    \bigoplus_{j=|b-b'|+1,\ \!{\rm by}\ \!2}^{p'-|p'-b-b'|-1}\!\!\ketw{\R_{\kappa p,j}^{\al,0}}
    \oplus  \bigoplus_{\beta}^{b+b'-p'-1}\!\ketw{\R_{\kappa p,p'}^{\al,\beta}}\Big\}\nn
 \ketw{a,b}\otimes\ketw{a',\kappa p'}&=& \bigoplus_{\beta}^{b-1}\Big\{
    \bigoplus_{i=|a-a'|+1,\ \!{\rm by}\ \!2}^{p-|p-a-a'|-1}\!\!\ketw{\R_{i,\kappa p'}^{0,\beta}}
    \oplus  \bigoplus_{\al}^{a+a'-p-1}\!\ketw{\R_{\kappa p,p'}^{\al,\beta}}\Big\}  \nn
 \ketw{a,b}\otimes\ketw{\kappa p,p'}&=&\bigoplus_{\al}^{a-1}\Big\{
  \bigoplus_{\beta}^{b-1}\ketw{\R_{\kappa p,p'}^{\al,\beta}}\Big\} 
\label{sup1}
\eea
and
\bea
 \ketw{a,b}\otimes\ketw{\R_{\kappa p,s}^{a',0}}
  &=&\bigoplus_{\al=|a-a'|+1,\ \!{\rm by}\ \!2}^{p-|p-a-a'|-1}
   \Big\{\bigoplus_{j=|b-s|+1,\ \!{\rm by}\ \!2}^{p'-|p'-b-s|-1}\ketw{\R_{\kappa p,j}^{\al,0}}
    \oplus \bigoplus_{\beta}^{b+s-p'-1}\ketw{\R_{\kappa p,p'}^{\al,\beta}}\Big\}  \nn
  &&\oplus  \bigoplus_{\al}^{a-a'-1}
   \Big\{\bigoplus_{j=|b-s|+1,\ \!{\rm by}\ \!2}^{p'-|p'-b-s|-1}2\ketw{\R_{\kappa p,j}^{\al,0}}
    \oplus \bigoplus_{\beta}^{b+s-p'-1}2\ketw{\R_{\kappa p,p'}^{\al,\beta}}\Big\}  \nn
        \nn
  &&\oplus  \bigoplus_{\al}^{a+a'-p-1}
   \Big\{\bigoplus_{j=|b-s|+1,\ \!{\rm by}\ \!2}^{p'-|p'-b-s|-1}2\ketw{\R_{(2\cdot\kappa)p,j}^{\al,0}}
    \oplus \bigoplus_{\beta}^{b+s-p'-1}2\ketw{\R_{(2\cdot\kappa)p,p'}^{\al,\beta}}\Big\} 
        \nn
 \ketw{a,b}\otimes\ketw{\R_{r,\kappa p'}^{0,b'}}
  &=&   \bigoplus_{\beta=|b-b'|+1,\ \!{\rm by}\ \!2}^{p'-|p'-b-b'|-1}
   \Big\{\bigoplus_{i=|a-r|+1,\ \!{\rm by}\ \!2}^{p-|p-a-r|-1}\ketw{\R_{i,\kappa p'}^{0,\beta}}
    \oplus \bigoplus_{\al}^{a+r-p-1}\ketw{\R_{\kappa p,p'}^{\al,\beta}}\Big\}  \nn
  &&\oplus  \bigoplus_{\beta}^{b-b'-1}
   \Big\{\bigoplus_{i=|a-r|+1,\ \!{\rm by}\ \!2}^{p-|p-a-r|-1}2\ketw{\R_{i,\kappa p'}^{0,\beta}}
    \oplus \bigoplus_{\al}^{a+r-p-1}2\ketw{\R_{\kappa p,p'}^{\al,\beta}}\Big\}  \nn
        \nn
  &&\oplus  \bigoplus_{\beta}^{b+b'-p'-1}
   \Big\{\bigoplus_{i=|a-r|+1,\ \!{\rm by}\ \!2}^{p-|p-a-r|-1}2\ketw{\R_{i,(2\cdot\kappa)p'}^{0,\beta}}
    \oplus \bigoplus_{\al}^{a+r-p-1}2\ketw{\R_{(2\cdot\kappa)p,p'}^{\al,\beta}}\Big\} 
\label{sup2}
\eea
and 
\bea
 \ketw{a,b}\otimes\ketw{\R_{\kappa p,p'}^{a',b'}}
  &=&\bigoplus_{\al=|a-a'|+1,\ \!{\rm by}\ \!2}^{p-|p-a-a'|-1}
   \Big\{\bigoplus_{\beta=|b-b'|+1,\ \!{\rm by}\ \!2}^{p'-|p'-b-b'|-1}
     \!\!\ketw{\R_{\kappa p,p'}^{\al,\beta}}\Big\}
   \nn
 &&\oplus\bigoplus_{\beta}^{b-b'-1}\Big\{
  \bigoplus_{\al=|a-a'|+1,\ \!{\rm by}\ \!2}^{p-|p-a-a'|-1}\!\!2\ketw{\R_{\kappa p,p'}^{\al,\beta}}\Big\}
  \oplus\bigoplus_{\al}^{a-a'-1}\Big\{
   \bigoplus_{\beta=|b-b'|+1,\ \!{\rm by}\ \!2}^{p'-|p'-b-b'|-1}\!\!2\ketw{\R_{\kappa p,p'}^{\al,\beta}}\Big\}
  \nn
 &&\oplus\bigoplus_{\al}^{a-a'-1}\Big\{\bigoplus_{\beta}^{b-b'-1}
   4\ketw{\R_{\kappa p,p'}^{\al,\beta}}\Big\}
  \oplus
    \bigoplus_{\al}^{a+a'-p-1}\Big\{\bigoplus_{\beta}^{b+b'-p'-1}
      4\ketw{\R_{\kappa p,p'}^{\al,\beta}}\Big\}
  \nn
&&\oplus\bigoplus_{\al}^{a+a'-p-1}\Big\{
  \bigoplus_{\beta=|b-b'|+1,\ \!{\rm by}\ \!2}^{p'-|p'-b-b'|-1}\!\!
    2\ketw{\R_{(2\cdot\kappa)p,p'}^{\al,\beta}}
    \oplus\bigoplus_{\beta}^{b-b'-1}4\ketw{\R_{(2\cdot\kappa)p,p'}^{\al,\beta}}\Big\}
      \nn
 &&\oplus\bigoplus_{\beta}^{b+b'-p'-1}\Big\{
  \bigoplus_{\al=|a-a'|+1,\ \!{\rm by}\ \!2}^{p-|p-a-a'|-1}\!\!2\ketw{\R_{(2\cdot\kappa)p,p'}^{\al,\beta}}
    \oplus\bigoplus_{\al}^{a-a'-1}4\ketw{\R_{(2\cdot\kappa)p,p'}^{\al,\beta}}\Big\}
\label{sup3}
\eea
The fusion of two ${\cal W}$-indecomposable rank-1 modules 
in $\ketw{\Jc^{\mathrm{out}}_{p,p'}}$ is given by
\bea
 \ketw{\kappa p,s}\otimes\ketw{\kappa'p,s'}&=&\bigoplus_{\al}^{p-1}\Big\{
  \!\bigoplus_{j=|s-s'|+1,\ \!{\rm by}\ \!2}^{p'-|p'-s-s'|-1}
  \!\!\!\ketw{\R_{(\kappa\cdot\kappa')p,j}^{\al,0}}
    \oplus\!\bigoplus_{\beta}^{s+s'-p'-1}
    \!\ketw{\R_{\kappa p,\kappa'p'}^{\al,\beta}}
    \Big\}   \nn
 \ketw{\kappa p,s}\otimes\ketw{r,\kappa'p'}&=&\bigoplus_{\al}^{r-1}
  \Big\{\bigoplus_{\beta}^{s-1}
     \ketw{\R_{\kappa p,\kappa'p'}^{\al,\beta}}\Big\}\nn
 \ketw{r,\kappa p'}\otimes\ketw{r',\kappa'p'}&=&\bigoplus_{\beta}^{p'-1}\Big\{
   \!\bigoplus_{j=|r-r'|+1,\ \!{\rm by}\ \!2}^{p-|p-r-r'|-1}
  \!\!\!\ketw{\R_{j,(\kappa\cdot\kappa')p'}^{0,\beta}}
  \oplus\!\bigoplus_{\al}^{r+r'-p-1}
    \!\ketw{\R_{\kappa p,\kappa'p'}^{\al,\beta}}
    \Big\}
\label{fus11}
\eea
The fusion of a ${\cal W}$-indecomposable rank-1 module in
$\ketw{\Jc^{\mathrm{out}}_{p,p'}}$ with a 
${\cal W}$-indecomposable rank-2 module is given by
\bea
 \ketw{\kappa p,s}\otimes\ketw{\R_{\kappa'p,s'}^{a,0}}
   &=&\!\bigoplus_{j=|s-s'|+1,\ \!{\rm by}\ \!2}^{p'-|p'-s-s'|-1}
  \!\!\Big\{\bigoplus_{\al}^{p-a-1}\!\!2\ketw{\R_{(\kappa\cdot\kappa')p,j}^{\al,0}}\oplus
    \bigoplus_{\al}^{a-1}2\ketw{\R_{(2\cdot\kappa\cdot\kappa')p,j}^{\al,0}}\Big\}  \nn
  &\oplus&\bigoplus_{\beta}^{s+s'-p'-1}
  \!\!\Big\{\bigoplus_{\al}^{p-a-1}\!\!2\ketw{\R_{\kappa p,\kappa'p'}^{\al,\beta}}\oplus
    \bigoplus_{\al}^{a-1}2\ketw{\R_{\kappa p,(2\cdot\kappa')p'}^{\al,\beta}}\Big\}  \nn
 \ketw{\kappa p,s}\otimes\ketw{\R_{r,\kappa'p'}^{0,b}}
  &=&\bigoplus_{\al}^{r-1}\Big\{
    \!\bigoplus_{\beta=|b-s|+1,\ \!{\rm by}\ \!2}^{p'-|p'-s-b|-1}
      \!\!\!\!\ketw{\R_{\kappa p,\kappa'p'}^{\al,\beta}}
     \oplus\!\bigoplus_{\beta}^{s-b-1}\!2\ketw{\R_{\kappa p,\kappa'p'}^{\al,\beta}}\oplus
     \!\!\!\bigoplus_{\beta}^{b+s-p'-1}
        \!\!\!2\ketw{\R_{\kappa p,(2\cdot\kappa')p'}^{\al,\beta}}\Big\}   \nn
 \ketw{r,\kappa p'}\otimes\ketw{\R_{\kappa'p,s}^{a,0}}
  &=&\bigoplus_{\beta}^{s-1}\Big\{
    \!\bigoplus_{\al=|a-r|+1,\ \!{\rm by}\ \!2}^{p-|p-r-a|-1}
      \!\!\!\!\ketw{\R_{\kappa p,\kappa'p'}^{\al,\beta}}
    \oplus\!\bigoplus_{\al}^{r-a-1}\!2\ketw{\R_{\kappa p,\kappa'p'}^{\al,\beta}}\oplus 
     \!\!\!\bigoplus_{\al}^{a+r-p-1}
        \!\!\!2\ketw{\R_{\kappa p,(2\cdot\kappa')p'}^{\al,\beta}}\Big\}   \nn
 \!\ketw{r,\kappa p'}\otimes\ketw{\R_{r',\kappa'p'}^{0,b}}
  &=&\!\bigoplus_{j=|r-r'|+1,\ \!{\rm by}\ \!2}^{p-|p-r-r'|-1}
  \!\!\Big\{\bigoplus_{\beta}^{p'-b-1}\!\!2\ketw{\R_{j,(\kappa\cdot\kappa')p'}^{0,\beta}}\oplus
    \bigoplus_{\beta}^{b-1}2\ketw{\R_{j,(2\cdot\kappa\cdot\kappa')p'}^{0,\beta}}\Big\}  \nn
  &\oplus&\bigoplus_{\al}^{r+r'-p-1}
  \!\!\Big\{\bigoplus_{\beta}^{p'-b-1}\!\!2\ketw{\R_{\kappa p,\kappa'p'}^{\al,\beta}}\oplus
    \bigoplus_{\beta}^{b-1}2\ketw{\R_{\kappa p,(2\cdot\kappa')p'}^{\al,\beta}}\Big\}   
\label{fus12}
\eea
The fusion of a ${\cal W}$-indecomposable rank-1 module in
$\ketw{\Jc^{\mathrm{out}}_{p,p'}}$ with a 
${\cal W}$-indecomposable rank-3 module is given by
\bea
 \ketw{\kappa p,s}\otimes\ketw{\R_{p,\kappa'p'}^{a,b}}
  \!&=&\!\!\!\bigoplus_{\al}^{p-a-1}\!\!\Big\{
     \!\bigoplus_{\beta=|b-s|+1,\ \!{\rm by}\ \!2}^{p'-|p'-s-b|-1}
      \!\!\!\!2\ketw{\R_{\kappa p,\kappa'p'}^{\al,\beta}}
   \oplus\!\bigoplus_{\beta}^{s-b-1}\!\!4\ketw{\R_{\kappa p,\kappa'p'}^{\al,\beta}}
     \oplus\!\!\!\!\bigoplus_{\beta}^{b+s-p'-1}
        \!\!\!\!4\ketw{\R_{\kappa p,(2\cdot\kappa')p'}^{\al,\beta}}\Big\}\nn
 \!&\oplus&\!\!\!\bigoplus_{\al}^{a-1}\!\Big\{
     \!\!\bigoplus_{\beta=|b-s|+1,\ \!{\rm by}\ \!2}^{p'-|p'-s-b|-1}
      \!\!\!\!2\ketw{\R_{\kappa p,(2\cdot\kappa')p'}^{\al,\beta}}
   \oplus\!\bigoplus_{\beta}^{s-b-1}\!\!4\ketw{\R_{\kappa p,(2\cdot\kappa')p'}^{\al,\beta}}
     \oplus\!\!\!\!\!\bigoplus_{\beta}^{b+s-p'-1}
        \!\!\!\!\!4\ketw{\R_{\kappa p,\kappa'p'}^{\al,\beta}}\Big\}  \nn
 \ketw{r,\kappa p'}\otimes\ketw{\R_{p,\kappa'p'}^{a,b}}
  \!&=&\!\!\!\bigoplus_{\beta}^{p'-b-1}\!\!\!\Big\{
     \!\bigoplus_{\al=|a-r|+1,\ \!{\rm by}\ \!2}^{p-|p-r-a|-1}
      \!\!\!\!2\ketw{\R_{\kappa p,\kappa'p'}^{\al,\beta}}
  \oplus\!\!\bigoplus_{\al}^{r-a-1}\!\!4\ketw{\R_{\kappa p,\kappa'p'}^{\al,\beta}}
     \oplus\!\!\!\!\bigoplus_{\al}^{a+r-p-1}
        \!\!\!\!4\ketw{\R_{\kappa p,(2\cdot\kappa')p'}^{\al,\beta}}\Big\}\nn
 \!&\oplus&\!\!\!\bigoplus_{\beta}^{b-1}\!\Big\{
     \!\bigoplus_{\al=|a-r|+1,\ \!{\rm by}\ \!2}^{p-|p-r-a|-1}
      \!\!\!\!2\ketw{\R_{\kappa p,(2\cdot\kappa')p'}^{\al,\beta}}
  \oplus\!\!\bigoplus_{\al}^{r-a-1}\!\!\!4\ketw{\R_{\kappa p,(2\cdot\kappa')p'}^{\al,\beta}}
     \oplus\!\!\!\!\!\bigoplus_{\al}^{a+r-p-1}
        \!\!\!\!\!4\ketw{\R_{\kappa p,\kappa'p'}^{\al,\beta}}\Big\}
     \nn
\label{fus13}
\eea
The fusion of two ${\cal W}$-indecomposable rank-2 modules is given by
\bea
 \ketw{\R_{\kappa p,s}^{a,0}}\otimes\ketw{\R_{\kappa'p,s'}^{a',0}}
  &=&\!\bigoplus_{j=|s-s'|+1,\ \!{\rm by}\ \!2}^{p'-|p'-s-s'|-1}
   \!\!\Big\{\bigoplus_{\al}^{p-|a-a'|-1}
     \!\!2\ketw{\R_{(\kappa\cdot\kappa')p,j}^{\al,0}}
     \oplus\!\bigoplus_{\al}^{|p-a-a'|-1}
     \!\!2\ketw{\R_{(\kappa\cdot\kappa')p,j}^{\al,0}}\nn
   &&\qquad\oplus\bigoplus_{\al}^{p-|p-a-a'|-1}
     \!\!\!2\ketw{\R_{(2\cdot\kappa\cdot\kappa')p,j}^{\al,0}}
     \oplus\bigoplus_{\al}^{|a-a'|-1}
     \!2\ketw{\R_{(2\cdot\kappa\cdot\kappa')p,j}^{\al,0}}\Big\}\nn
&\oplus&\bigoplus_{\beta}^{s+s'-p'-1}   \!\!\Big\{\bigoplus_{\al}^{p-|a-a'|-1}
     \!\!2\ketw{\R_{\kappa p,\kappa'p'}^{\al,\beta}}
     \oplus\bigoplus_{\al}^{|p-a-a'|-1}
     \!\!2\ketw{\R_{\kappa p,\kappa'p'}^{\al,\beta}}\nn
   &&\qquad\oplus\bigoplus_{\al}^{p-|p-a-a'|-1}
     \!\!\!2\ketw{\R_{\kappa p,(2\cdot\kappa')p'}^{\al,\beta}}
     \oplus\bigoplus_{\al}^{|a-a'|-1}
     \!2\ketw{\R_{\kappa p,(2\cdot\kappa')p'}^{\al,\beta}}\Big\}\nn
 \ketw{\R_{\kappa p,s}^{a,0}}\otimes\ketw{\R_{r,\kappa'p'}^{0,b}}
  &=&
   \!\!\bigoplus_{\al=|a-r|+1,\ \!{\rm by}\ \!2}^{p-|p-r-a|-1}
   \!\Big\{\bigoplus_{\beta=|b-s|+1,\ \!{\rm by}\ \!2}^{p'-|p'-s-b|-1}
    \!\!\!\!\ketw{\R_{\kappa p,\kappa'p'}^{\al,\beta}}\Big\}
    \nn
 &\oplus&  
     \!\!\bigoplus_{\al=|a-r|+1,\ \!{\rm by}\ \!2}^{p-|p-r-a|-1}
   \!\Big\{  \bigoplus_{\beta}^{s-b-1}2\ketw{\R_{\kappa p,\kappa'p'}^{\al,\beta}}\Big\}
    \oplus\bigoplus_{\beta=|b-s|+1,\ \!{\rm by}\ \!2}^{p'-|p'-s-b|-1}
    \!\Big\{  \bigoplus_{\al}^{r-a-1}2\ketw{\R_{\kappa p,\kappa'p'}^{\al,\beta}}\Big\}
    \nn
  &\oplus&\!\bigoplus_{\al}^{r-a-1}\!\Big\{ 
     \bigoplus_{\beta}^{s-b-1}4\ketw{\R_{\kappa p,\kappa'p'}^{\al,\beta}}\Big\}
      \oplus\!\bigoplus_{\al}^{a+r-p-1}\!\Big\{
     \bigoplus_{\beta}^{b+s-p'-1}\!4\ketw{\R_{\kappa p,\kappa'p'}^{\al,\beta}}\Big\}
     \nn
 &\oplus&\!\bigoplus_{\al}^{a+r-p-1}\!\Big\{
   \bigoplus_{\beta=|b-s|+1,\ \!{\rm by}\ \!2}^{p'-|p'-s-b|-1}
     \!\!2\ketw{\R_{\kappa p,(2\cdot\kappa')p'}^{\al,\beta}}\oplus
      \bigoplus_{\beta}^{s-b-1}\!4\ketw{\R_{\kappa p,(2\cdot\kappa')p'}^{\al,\beta}}\Big\}
      \nn
 &\oplus&\!\bigoplus_{\beta}^{b+s-p'-1}\!\Big\{
  \bigoplus_{\al=|a-r|+1,\ \!{\rm by}\ \!2}^{p-|p-r-a|-1}
    \!\!2\ketw{\R_{\kappa p,(2\cdot\kappa')p'}^{\al,\beta}}\oplus
     \bigoplus_{\beta}^{r-a-1}\!4\ketw{\R_{\kappa p,(2\cdot\kappa')p'}^{\al,\beta}}\Big\}
     \nn
     \nn
 \ketw{\R_{r,\kappa p'}^{0,b}}\otimes\ketw{\R_{r',\kappa'p'}^{0,b'}}
  &=&\!\!\bigoplus_{j=|r-r'|+1,\ \!{\rm by}\ \!2}^{p-|p-r-r'|-1}
   \!\!\Big\{\bigoplus_{\beta}^{p'-|b-b'|-1}
     \!\!2\ketw{\R_{j,(\kappa\cdot\kappa')p'}^{0,\beta}}
     \oplus\bigoplus_{\beta}^{|p'-b-b'|-1}
     \!\!2\ketw{\R_{j,(\kappa\cdot\kappa')p'}^{0,\beta}}\nn
   &&\qquad\oplus\bigoplus_{\beta}^{p'-|p'-b-b'|-1}
     \!\!\!2\ketw{\R_{j,(2\cdot\kappa\cdot\kappa')p'}^{0,\beta}}
     \oplus\bigoplus_{\beta}^{|b-b'|-1}
     \!2\ketw{\R_{j,(2\cdot\kappa\cdot\kappa')p'}^{0,\beta}}\Big\}\nn
&\oplus&\bigoplus_{\al}^{r+r'-p-1}   \!\Big\{\bigoplus_{\beta}^{p'-|b-b'|-1}
     \!\!2\ketw{\R_{\kappa p,\kappa'p'}^{\al,\beta}}
     \oplus\!\bigoplus_{\beta}^{|p'-b-b'|-1}
     \!\!2\ketw{\R_{\kappa p,\kappa'p'}^{\al,\beta}}\nn
   &&\qquad\oplus\!\bigoplus_{\beta}^{p'-|p'-b-b'|-1}
     \!\!\!2\ketw{\R_{\kappa p,(2\cdot\kappa')p'}^{\al,\beta}}
     \oplus\bigoplus_{\beta}^{|b-b'|-1}
     \!2\ketw{\R_{\kappa p,(2\cdot\kappa')p'}^{\al,\beta}}\Big\}  
\label{fus22}
\eea
The fusion of a ${\cal W}$-indecomposable rank-2 module with a 
${\cal W}$-indecomposable rank-3 module is given by
\bea
 \ketw{\R_{\kappa p,s}^{a,0}}\otimes\ketw{\R_{p,\kappa'p'}^{a',b'}}
  &=&\!\!\bigoplus_{\beta=|b'-s|+1,\ \!{\rm by}\ \!2}^{p'-|p'-s-b'|-1}\!\!\Big\{
   \bigoplus_{\al}^{p-|a-a'|-1}\!\!\!2\ketw{\R_{\kappa p,\kappa'p'}^{\al,\beta}}
   \oplus\!\bigoplus_{\al}^{|p-a-a'|-1}\!\!\!2\ketw{\R_{\kappa p,\kappa'p'}^{\al,\beta}}\Big\}
  \nn
  &\oplus&\bigoplus_{\beta}^{s-b'-1}\!\Big\{
   \bigoplus_{\al}^{p-|a-a'|-1}\!\!\!4\ketw{\R_{\kappa p,\kappa'p'}^{\al,\beta}}
   \oplus\!\bigoplus_{\al}^{|p-a-a'|-1}\!\!\!4\ketw{\R_{\kappa p,\kappa'p'}^{\al,\beta}}\Big\}
  \nn
  &\oplus&\bigoplus_{\beta}^{b'+s-p'-1}\!\Big\{
   \bigoplus_{\al}^{p-|p-a-a'|-1}\!\!\!4\ketw{\R_{\kappa p,\kappa'p'}^{\al,\beta}}\oplus
   \bigoplus_{\al}^{|a-a'|-1}\!4\ketw{\R_{\kappa p,\kappa'p'}^{\al,\beta}}\Big\}
 \nn
 &\oplus&\!\!\bigoplus_{\beta=|b'-s|+1,\ \!{\rm by}\ \!2}^{p'-|p'-s-b'|-1}\!\!\Big\{
   \bigoplus_{\al}^{p-|p-a-a'|-1}\!\!\!2\ketw{\R_{\kappa p,(2\cdot\kappa')p'}^{\al,\beta}}
   \oplus\bigoplus_{\al}^{|a-a'|-1}\!2\ketw{\R_{\kappa p,(2\cdot\kappa')p'}^{\al,\beta}}\Big\}
 \nn
 &\oplus&\bigoplus_{\beta}^{s-b'-1}\!\Big\{
   \bigoplus_{\al}^{p-|p-a-a'|-1}\!\!\!4\ketw{\R_{\kappa p,(2\cdot\kappa')p'}^{\al,\beta}}\oplus
   \bigoplus_{\al}^{|a-a'|-1}\!4\ketw{\R_{\kappa p,(2\cdot\kappa')p'}^{\al,\beta}}\Big\}
 \nn
 &\oplus&\bigoplus_{\beta}^{b'+s-p'-1}\!\Big\{
   \bigoplus_{\al}^{p-|a-a'|-1}\!\!\!4\ketw{\R_{\kappa p,(2\cdot\kappa')p'}^{\al,\beta}}
   \oplus\!\bigoplus_{\al}^{|p-a-a'|-1}\!\!\!4\ketw{\R_{\kappa p,(2\cdot\kappa')p'}^{\al,\beta}}\Big\} 
 \nn 
 \ketw{\R_{r,\kappa p'}^{0,b}}\otimes\ketw{\R_{p,\kappa'p'}^{a',b'}}
  &=&\!\!\bigoplus_{\al=|a'-r|+1,\ \!{\rm by}\ \!2}^{p-|p-r-a'|-1}\!\!\Big\{
   \bigoplus_{\beta}^{p'-|b-b'|-1}\!\!\!2\ketw{\R_{\kappa p,\kappa'p'}^{\al,\beta}}
   \oplus\!\bigoplus_{\beta}^{|p'-b-b'|-1}\!\!\!2\ketw{\R_{\kappa p,\kappa'p'}^{\al,\beta}}\Big\}
  \nn
  &\oplus&\bigoplus_{\al}^{r-a'-1}\!\Big\{
   \bigoplus_{\beta}^{p'-|b-b'|-1}\!\!\!4\ketw{\R_{\kappa p,\kappa'p'}^{\al,\beta}}
   \oplus\!\bigoplus_{\beta}^{|p'-b-b'|-1}\!\!\!4\ketw{\R_{\kappa p,\kappa'p'}^{\al,\beta}}\Big\}
  \nn
  &\oplus&\bigoplus_{\al}^{a'+r-p-1}\!\Big\{
   \bigoplus_{\beta}^{p'-|p'-b-b'|-1}\!\!\!4\ketw{\R_{\kappa p,\kappa'p'}^{\al,\beta}}\oplus
   \bigoplus_{\beta}^{|b-b'|-1}\!4\ketw{\R_{\kappa p,\kappa'p'}^{\al,\beta}}\Big\}
 \nn
 &\oplus&\!\!\bigoplus_{\al=|a'-r|+1,\ \!{\rm by}\ \!2}^{p-|p-r-a'|-1}\!\!\Big\{
   \bigoplus_{\beta}^{p'-|p'-b-b'|-1}\!\!\!2\ketw{\R_{\kappa p,(2\cdot\kappa')p'}^{\al,\beta}}
   \oplus\bigoplus_{\beta}^{|b-b'|-1}\!2\ketw{\R_{\kappa p,(2\cdot\kappa')p'}^{\al,\beta}}\Big\}
 \nn
 &\oplus&\bigoplus_{\al}^{r-a'-1}\!\Big\{
   \bigoplus_{\beta}^{p'-|p'-b-b'|-1}\!\!\!4\ketw{\R_{\kappa p,(2\cdot\kappa')p'}^{\al,\beta}}\oplus
   \bigoplus_{\beta}^{|b-b'|-1}\!4\ketw{\R_{\kappa p,(2\cdot\kappa')p'}^{\al,\beta}}\Big\}
 \nn
 &\oplus&\bigoplus_{\al}^{a'+r-p-1}\!\Big\{
   \bigoplus_{\beta}^{p'-|b-b'|-1}\!\!\!4\ketw{\R_{\kappa p,(2\cdot\kappa')p'}^{\al,\beta}}
   \oplus\!\bigoplus_{\beta}^{|p'-b-b'|-1}\!\!\!4\ketw{\R_{\kappa p,(2\cdot\kappa')p'}^{\al,\beta}}\Big\}  
\label{fus23}
\eea
Finally, the fusion of two ${\cal W}$-indecomposable rank-3 modules is given by
\bea
 \ketw{\R_{\kappa p,p'}^{a,b}}\otimes\ketw{\R_{p,\kappa'p'}^{a',b'}}
 &=&\!\bigoplus_{\al}^{p-|a-a'|-1}\!\!\Big\{
  \bigoplus_{\beta}^{p'-|b-b'|-1}\!\!\!4\ketw{\R_{\kappa p,\kappa'p'}^{\al,\beta}}\Big\}
   \oplus\!\!\bigoplus_{\al}^{|p-a-a'|-1}\!\!\Big\{
  \bigoplus_{\beta}^{|p'-b-b'|-1}\!\!\!4\ketw{\R_{\kappa p,\kappa'p'}^{\al,\beta}}\Big\}
 \nn
 &\oplus&\!\bigoplus_{\al}^{p-|a-a'|-1}\!\!\Big\{
  \bigoplus_{\beta}^{|p'-b-b'|-1}\!\!\!4\ketw{\R_{\kappa p,\kappa'p'}^{\al,\beta}}\Big\}
   \oplus\!\!\bigoplus_{\al}^{|p-a-a'|-1}\!\!\Big\{
  \bigoplus_{\beta}^{p'-|b-b'|-1}\!\!\!4\ketw{\R_{\kappa p,\kappa'p'}^{\al,\beta}}\Big\}
 \nn
 &\oplus&\!\bigoplus_{\al}^{p-|p-a-a'|-1}\!\!\Big\{
  \bigoplus_{\beta}^{p'-|p'-b-b'|-1}\!\!\!4\ketw{\R_{\kappa p,\kappa'p'}^{\al,\beta}}\Big\}
   \oplus\!\!\bigoplus_{\al}^{|a-a'|-1}\!\!\Big\{
  \bigoplus_{\beta}^{|b-b'|-1}\!\!\!4\ketw{\R_{\kappa p,\kappa'p'}^{\al,\beta}}\Big\}
 \nn
 &\oplus&\!\bigoplus_{\al}^{p-|p-a-a'|-1}\!\!\Big\{
  \bigoplus_{\beta}^{|b-b'|-1}\!\!\!4\ketw{\R_{\kappa p,\kappa'p'}^{\al,\beta}}\Big\}
   \oplus\!\!\bigoplus_{\al}^{|a-a'|-1}\!\!\Big\{
  \bigoplus_{\beta}^{p'-|p'-b-b'|-1}\!\!\!4\ketw{\R_{\kappa p,\kappa'p'}^{\al,\beta}}\Big\}
 \nn
 &\oplus&\!\bigoplus_{\al}^{p-|a-a'|-1}\!\!\Big\{
  \bigoplus_{\beta}^{p'-|p'-b-b'|-1}\!\!\!4\ketw{\R_{\kappa p,(2\cdot\kappa')p'}^{\al,\beta}}
  \oplus\!\!\bigoplus_{\beta}^{|b-b'|-1}\!\!\!4\ketw{\R_{\kappa p,(2\cdot\kappa')p'}^{\al,\beta}}\Big\}
 \nn
 &\oplus&\!\bigoplus_{\al}^{|p-a-a'|-1}\!\!\Big\{
    \bigoplus_{\beta}^{p'-|p'-b-b'|-1}\!\!\!4\ketw{\R_{\kappa p,(2\cdot\kappa')p'}^{\al,\beta}}
  \oplus\!\!\bigoplus_{\beta}^{|b-b'|-1}\!\!\!4\ketw{\R_{\kappa p,(2\cdot\kappa')p'}^{\al,\beta}}\Big\}
 \nn
 &\oplus&\!\bigoplus_{\beta}^{p'-|b-b'|-1}\!\!\Big\{
  \bigoplus_{\al}^{p-|p-a-a'|-1}\!\!\!4\ketw{\R_{\kappa p,(2\cdot\kappa')p'}^{\al,\beta}}
  \oplus\!\!\bigoplus_{\al}^{|a-a'|-1}\!\!\!4\ketw{\R_{\kappa p,(2\cdot\kappa')p'}^{\al,\beta}}\Big\}
 \nn
 &\oplus&\!\bigoplus_{\beta}^{|p'-b-b'|-1}\!\!\Big\{
    \bigoplus_{\al}^{p-|p-a-a'|-1}\!\!\!4\ketw{\R_{\kappa p,(2\cdot\kappa')p'}^{\al,\beta}}
  \oplus\!\!\bigoplus_{\al}^{|a-a'|-1}\!\!\!4\ketw{\R_{\kappa p,(2\cdot\kappa')p'}^{\al,\beta}}\Big\}
\label{fus33}
\eea

\subsection{Some fusion evaluations}
\label{AppFusB}

The applications of $\Kc$, $\Lc$, $\Lc^2$, $\Mc$ and $\Mc^2$ appearing in
(\ref{fusconj2}) read
\bea
 \Kc\big[\ketw{a,b}\otimes\ketw{a',b'}\big]&=&
   \bigoplus_{i=|a-a'|+1,\ \!{\rm by}\ \!2}^{p-|p-a-a'|-1}
     \ \bigoplus_{j=|b-b'|+1,\ \!{\rm by}\ \!2}^{p'-|p'-b-b'|-1}
  \!\!\!\Wc(\D_{3p-i,j})\oplus\!
  \bigoplus_{\al}^{a+a'-p-1}\!
   \ \bigoplus_{j=|b-b'|+1,\ \!{\rm by}\ \!2}^{p'-|p'-b-b'|-1}\ketw{\R_{2p,j}^{\al,0}}\nn
 &&\oplus\!\bigoplus_{\beta}^{b+b'-p'-1}\!
   \ \bigoplus_{i=|a-a'|+1,\ \!{\rm by}\ \!2}^{p-|p-a-a'|-1}\ketw{\R_{i,2p'}^{0,\beta}}
  \oplus\bigoplus_{\al}^{a+a'-p-1}
    \ \bigoplus_{\beta}^{b+b'-p'-1}\!\ketw{\R_{2p,p'}^{\al,\beta}}
  \nn
 \Lc\big[\ketw{a,b}\otimes\ketw{a',b'}\big]&=&
   \bigoplus_{i=|a-a'|+1,\ \!{\rm by}\ \!2}^{p-|p-a-a'|-1}
     \ \bigoplus_{j=|b-b'|+1,\ \!{\rm by}\ \!2}^{p'-|p'-b-b'|-1}
  \!\!\!\Wc(\D_{2p-i,j})\oplus\!
  \bigoplus_{\al}^{a+a'-p-1}\!
   \ \bigoplus_{j=|b-b'|+1,\ \!{\rm by}\ \!2}^{p'-|p'-b-b'|-1}\ketw{\R_{p,j}^{\al,0}}\nn
 &&\oplus\!\bigoplus_{\beta}^{b+b'-p'-1}\!
   \ \bigoplus_{i=|a-a'|+1,\ \!{\rm by}\ \!2}^{p-|p-a-a'|-1}\ketw{\R_{i,p'}^{0,\beta}}
  \oplus\bigoplus_{\al}^{a+a'-p-1}
    \ \bigoplus_{\beta}^{b+b'-p'-1}\!\ketw{\R_{p,p'}^{\al,\beta}}
  \nn
 \Lc^2\big[\ketw{a,b}\otimes\ketw{a',b'}\big]&=&
   \bigoplus_{i=|a-a'|+1,\ \!{\rm by}\ \!2}^{p-|p-a-a'|-1}
     \ \bigoplus_{j=|b-b'|+1,\ \!{\rm by}\ \!2}^{p'-|p'-b-b'|-1}
  \!\!\!\ketwast{i,j}\oplus\!
  \bigoplus_{\al}^{a+a'-p-1}\!
   \ \bigoplus_{j=|b-b'|+1,\ \!{\rm by}\ \!2}^{p'-|p'-b-b'|-1}\ketw{\R_{p,j}^{\al,0}}\nn
 &&\oplus\!\bigoplus_{\beta}^{b+b'-p'-1}\!
   \ \bigoplus_{i=|a-a'|+1,\ \!{\rm by}\ \!2}^{p-|p-a-a'|-1}\ketw{\R_{i,p'}^{0,\beta}}
  \oplus\bigoplus_{\al}^{a+a'-p-1}
    \ \bigoplus_{\beta}^{b+b'-p'-1}\!\ketw{\R_{p,p'}^{\al,\beta}}
  \nn
 \Mc\big[\ketw{a,b}\otimes\ketw{a',b'}\big]&=&
   \bigoplus_{i=|a-a'|+1,\ \!{\rm by}\ \!2}^{p-|p-a-a'|-1}
     \ \bigoplus_{j=|b-b'|+1,\ \!{\rm by}\ \!2}^{p'-|p'-b-b'|-1}
  \!\!\!\Wc(\D_{p-i,j})
  \nn
 \Mc^2\big[\ketw{a,b}\otimes\ketw{a',b'}\big]&=&
   \bigoplus_{i=|a-a'|+1,\ \!{\rm by}\ \!2}^{p-|p-a-a'|-1}
     \ \bigoplus_{j=|b-b'|+1,\ \!{\rm by}\ \!2}^{p'-|p'-b-b'|-1}
  \!\!\!\Wc(\D_{i,j})
\eea

\subsection{Basic fusion rules}
\label{AppBasic}

For $p>2$, the basic fusion rules of $\mathrm{Conj}[{\cal WLM}(p,p')]$ (\ref{Conj6}) are
summarized in Figure~\ref{pgeq3}, where 
$\ketw{3,1}=\Wc(\D_{3,1})\in\ketw{\Jc^{\mathrm{out}}_{3,p'}}$ for $p=3$.
The basic fusion rules for $p=2$ are given in Figure~\ref{peq2},
where $\ketw{2,1}=\Wc(\D_{2,1})\in\ketw{\Jc^{\mathrm{out}}_{2,p'}}$ 
and, for $p'=3$, $\ketw{1,3}=\Wc(\D_{1,3})\in\ketw{\Jc^{\mathrm{out}}_{2,3}}$.
\begin{figure}
\scriptsize
$$
\renewcommand{\arraystretch}{1.5}
\begin{array}{c||cccccc}
 \otimes &\ketw{1,1}&\ketw{2,1}&\ketw{1,2}&\Wc(\D_{p-1,1})&\Wc(\D_{2p-1,1})&\Wc(\D_{3p-1,1})
\\[4pt]
\hline \hline
\rule{0pt}{14pt}
 \ketw{1,1}
  &\ketw{1,1}&\ketw{2,1}&\ketw{1,2}&\Wc(\D_{p-1,1})&\Wc(\D_{2p-1,1})&\Wc(\D_{3p-1,1})
\\[4pt]
 \ketw{2,1}
  &\ketw{2,1}&\ketw{1,1}\oplus\ketw{3,1}&\ketw{2,2}
      &\Wc(\D_{p-2,1})&\Wc(\D_{2p-2,1})&\Wc(\D_{3p-2,1})
\\[4pt]
 \ketw{1,2}
  &\ketw{1,2}&\ketw{2,2}&\ketw{1,1}\oplus\ketw{1,3}
     &\Wc(\D_{p-1,2})&\Wc(\D_{2p-1,2})&\Wc(\D_{3p-1,2})
\\[4pt]
 \Wc(\D_{p-1,1})
  &\Wc(\D_{p-1,1})&\Wc(\D_{p-2,1})&\Wc(\D_{p-1,2})
      &\Wc(\D_{1,1})&0&0
\\[4pt]
 \Wc(\D_{2p-1,1})
  &\Wc(\D_{2p-1,1})&\Wc(\D_{2p-2,1})&\Wc(\D_{2p-1,2})
      &0&\ketwast{1,1}&\Wc(\D_{3p-1,1})
\\[4pt]
 \Wc(\D_{3p-1,1})
  &\Wc(\D_{3p-1,1})&\Wc(\D_{3p-2,1})&\Wc(\D_{3p-1,2})
      &0&\Wc(\D_{3p-1,1})&\ketwast{1,1}
\end{array}
$$
\caption{Cayley table of the basic fusion rules for $\mathrm{Conj}[{\cal WLM}(p,p')]$ for $p>2$.}
\label{pgeq3}
\end{figure}
\begin{figure}
\scriptsize
$$
\renewcommand{\arraystretch}{1.5}
\begin{array}{c||cccccc}
 \otimes &\ketw{1,1}&\Wc(\D_{2,1})&\ketw{1,2}&\Wc(\D_{1,1})&\Wc(\D_{3,1})&\Wc(\D_{5,1})
\\[4pt]
\hline \hline
\rule{0pt}{14pt}
 \ketw{1,1}
  &\ketw{1,1}&\Wc(\D_{2,1})&\ketw{1,2}&\Wc(\D_{1,1})&\Wc(\D_{3,1})&\Wc(\D_{5,1})
\\[4pt]
 \Wc(\D_{2,1})
  &\Wc(\D_{2,1})&\ketw{\R_{2,1}^{1,0}}&\Wc(\D_{2,2})
      &0&\Wc(\D_{2,1})&\Wc(\D_{4,1})
\\[4pt]
 \ketw{1,2}
  &\ketw{1,2}&\Wc(\D_{2,2})&\ketw{1,1}\oplus\ketw{1,3}
     &\Wc(\D_{1,2})&\Wc(\D_{3,2})&\Wc(\D_{5,2})
\\[4pt]
 \Wc(\D_{1,1})
  &\Wc(\D_{1,1})&0&\Wc(\D_{1,2})
      &\Wc(\D_{1,1})&0&0
\\[4pt]
 \Wc(\D_{3,1})
  &\Wc(\D_{3,1})&\Wc(\D_{2,1})&\Wc(\D_{3,2})
      &0&\ketwast{1,1}&\Wc(\D_{5,1})
\\[4pt]
 \Wc(\D_{5,1})
  &\Wc(\D_{5,1})&\Wc(\D_{4,1})&\Wc(\D_{5,2})
      &0&\Wc(\D_{5,1})&\ketwast{1,1}
\end{array}
$$
\caption{Cayley table of the basic fusion rules for $\mathrm{Conj}[{\cal WLM}(2,p')]$.}
\label{peq2}
\end{figure}
%
%


\end{document}